\documentclass[
aps,
amsmath,
amssymb,
preprint
]{revtex4-1}

\usepackage{graphicx}
\usepackage{epstopdf}
\usepackage{epsfig}
\usepackage{subfigure}
\usepackage{dcolumn}
\usepackage{bm}
\usepackage{array}
\newcolumntype{P}[1]{>{\centering\arraybackslash}p{#1}}
\newcolumntype{M}[1]{>{\centering\arraybackslash}m{#1}}
\newcommand\plotsw{18} 

\begin{document}


\title{Experimental investigation of the wake behind a rotating sphere}



\author{M. Skarysz}

\altaffiliation[Present address: ]{Department of Aeronautical and Automotive Engineering,\\ Loughborough University, Loughborough LE11 3TU, United Kingdom.}
\email[\\E-mail address: ]{M.Skarysz@lboro.ac.uk (M. Skarysz).}

\affiliation{Institute of Aeronautics and Applied Mechanics, Warsaw University of Technology, Nowowiejska 24, 00-665 Warsaw, Poland}

\author{S. Goujon-Durand}
\author{J. E. Wesfreid}
\email[E-mail address: ]{wesfreid@espci.fr (J. E. Wesfreid)}

\affiliation{Laboratoire de Physique et M\'{e}canique des Millieux H\'{e}t\'{e}rog\`{e}ne (PMMH), UMR 7636  ESPCI-CNRS-UPMC-UPD, Ecole Superieure de Physique et de Chimie Industrielles de Ville de Paris, 10, rue Vauquelin, 75231 Paris CEDEX 05, France}

\author{J. Rokicki}
\affiliation{Institute of Aeronautics and Applied Mechanics, Warsaw University of Technology, Nowowiejska 24, 00-665 Warsaw, Poland}


\date{\today}

\begin{abstract}
The wake behind a sphere, rotating about an axis aligned with the streamwise direction, has been experimentally investigated in a low-velocity water tunnel using LIF visualizations and PIV measurements. The measurements focused on the evolution of the flow regimes that appear depending on two control parameters, namely the Reynolds number $Re$ and the dimensionless rotation or swirl rate $\Omega$, which is the ratio of the maximum azimuthal velocity of the body to the free stream velocity. In the present investigation, we cover the range of $Re$ smaller than 400 and $\Omega$ from 0 and 4. Different wakes regimes such as an axisymmetric flow, a \textit{low helical} state and a \textit{high helical} mode are represented in the ($Re$, $\Omega$) parameter plane.
\end{abstract}

\pacs{}

\maketitle

\section{Introduction}
The wake and its instabilities behind different 3D static bodies, e.g., sphere, disks, cube, bullets, etc. were considered in the numerical as well as in the experimental studies \cite{Bobinski,Chrust,Gumowski,JohnsonPatel,Klotz,Szaltys}. The flow instabilities of a viscous fluid past a rotating sphere may be considered as a simplified case of a more general family of flows around immersed rotating bodies \cite{JimenezJFS,JimenezJFM}. Therefore a rotation of sphere remains a main objective of the present paper, as it will allow understanding the effect of rotation for more complicated flows (e.g., behind wind turbines and propellers \cite{felli,Iungo}).

The direction of rotation has a significant influence on the characteristics of the flow over a sphere. Two particular directions of rotation can be specified. The first one is orthogonal to the streamwise direction. In this case, research focuses on the so-called Magnus or Robins effect (Barkla and Auchterlone \citep{BarklaAuch}) and the determination of the lift and the drag forces. The other one is the rotation of the sphere parallel to the streamwise direction. Most works (Kim and Choi \cite{KimChoi}; Pier \cite{Pier}; Poon \textit{et al.}  \cite{Poon}) focus on forces (lift and drag) acting on the rotating sphere depending on values of the free stream velocity and the rotational speed. However, this research was limited only to the numerical approach. The influence of rotation direction on the wake was presented by Poon \textit{et al.} \cite{Poon}. Simulations were performed for the Reynolds number 100, 250 and 300 and for non-dimensional rotation rates $\Omega=0.05$--$1$ and for different angles of rotation (rotation rate $\Omega$ is defined as a ratio of the maximum azimuthal velocity on the rotating body and the free stream velocity). In particular, it was investigated for which parameters the flow becomes unstable. In contrast, the present paper is limited to the experimental approach and to the case of sphere rotating about an axis aligned with the streamwise direction.

Flow instabilities and bifurcations in the wake behind a static sphere ($\Omega=0$) are at present well recognized. The first three regimes are: (i) the axisymmetric basic flow, (ii) the planar symmetric flow with two steady counter rotating streamwise vortices \textit{2CRV} and  (iii) the unsteady periodic hairpin shedding \textit{HS}, which preserves the symmetry plane, preceded by peristaltic oscillations of the \textit{2CRV} \cite{Gumowski}. The bifurcations appear for the critical Reynolds numbers $Re_{1}=212$ and $Re_{2}=272$ \cite{JohnsonPatel,Gumowski,Szaltys}. For the rotating sphere, the flow regimes change significantly (as documented by the numerical results). Kim and Choi \cite{KimChoi} shows the evolution of vorticity structures as a function of $\Omega$ for specific Reynolds numbers, namely $100$, $250$ and $300$. In particular, for the steady planar symmetric flow with two counter-rotating vortices, it is shown that the sphere's rotation increases the intensity of one longitudinal vortex while the other one is weakened. Kim and Choi \cite{KimChoi} have determined the revolution frequency of the vortical structure as a function of the rotation rate $\Omega$ and of the Reynolds number. In a more comprehensive study Pier \cite{Pier} has defined the following flow regimes (i) the axisymmetric steady base flow (whether stable or unstable), (ii) the low-frequency periodic helical, (iii) the quasi-periodic shedding and (iv) the high-frequency periodic helical.

The main purpose of the present work is the experimental study of this flow (never carried out earlier) and in particular the investigation of the influence of the streamwise rotation rate $\Omega$ and the Reynolds number $Re$ on the wake flow past a sphere. We also study the existence of different flow regimes as well as the bifurcation thresholds and compare them with numerical results. We discovered new effects such as the nonlinear modification of the homogeneous component of axial vorticity with the increasing rotation rate.

Preliminary work was published in a conference proceedings \cite{Skarysz} but with incomplete results for the frequency behaviour and only at the Reynolds number $Re=250$ and $\Omega$ in the range of $0-1$ which are in extenso presented here.

The reader should note that the design of the experiment with a rotating body is quite a difficult task as the electric motor and the support system located upstream of the sphere should introduce only a minimal perturbation into the incoming flow. This is further explained in the text.

\section{Experimental set-up}

\begin{figure}
\centerline{\includegraphics[width=20 pc]{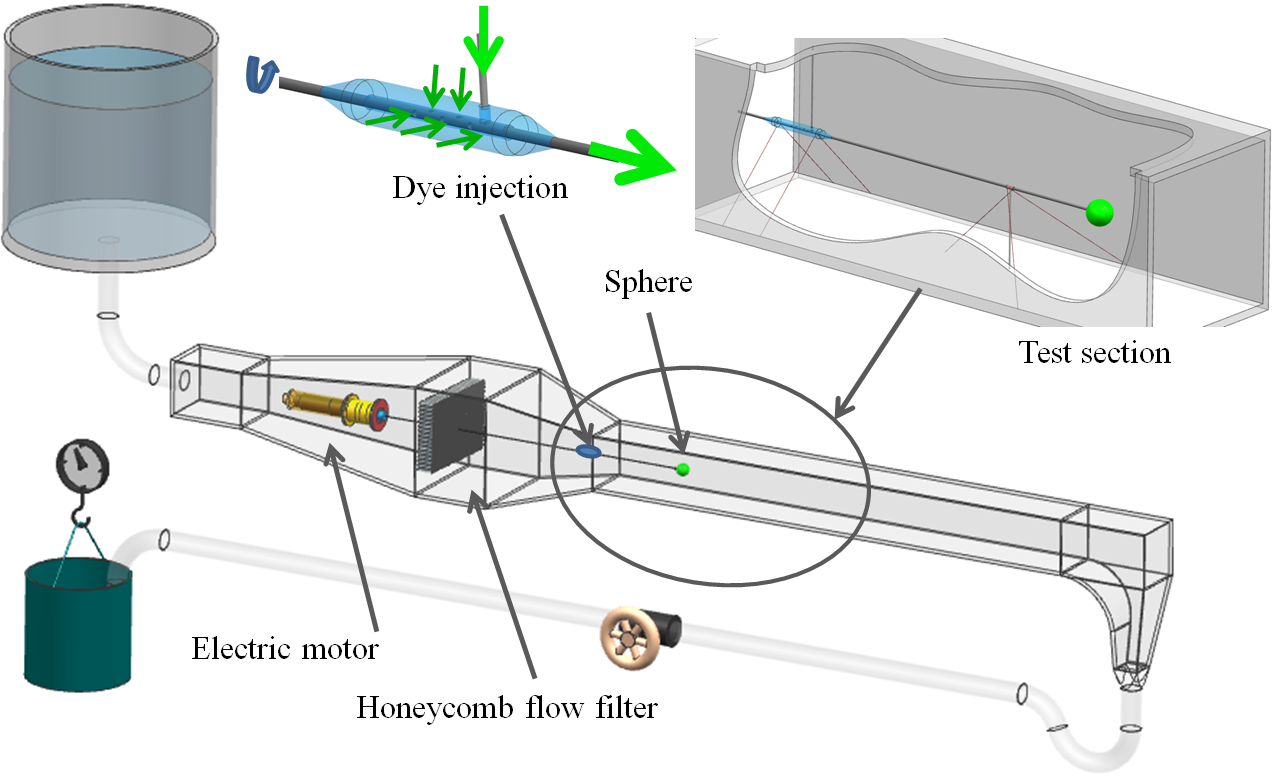}}
\caption{Schematic view of the experimental set-up. The water channel and the detailed view of test section showing the dye injection system and support. The sphere axis is supported by a group of $74\rm{\mu m}$ nylon wires with the closest wire located $5\rm{cm}$ upstream from the sphere.}
\label{ex}
\end{figure}

\begin{figure}
\centerline{\includegraphics[width=15 pc]{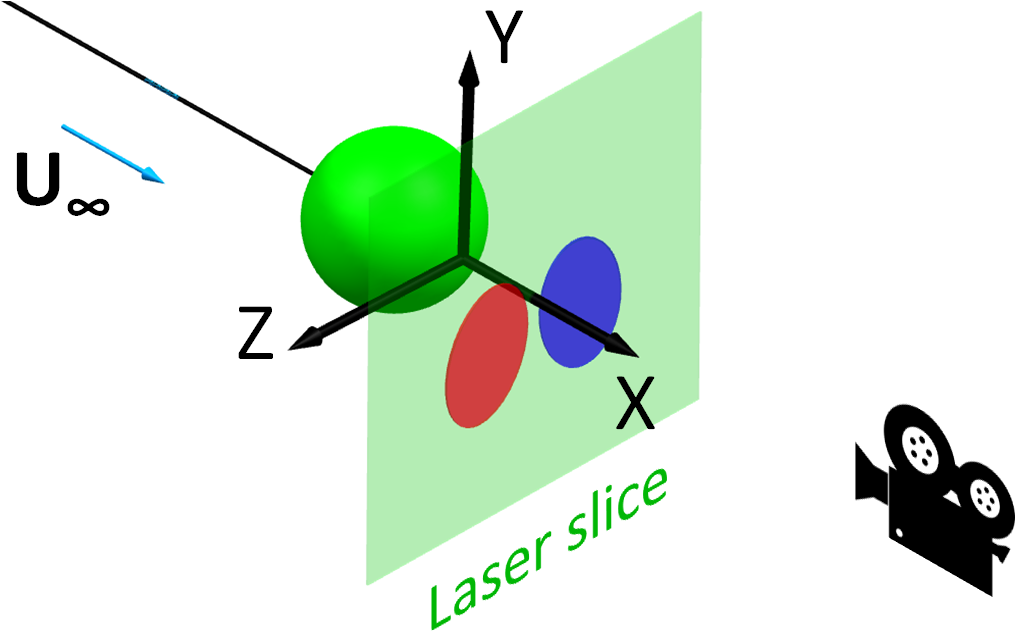}}
\caption{PIV laser plane at a given $x$ position behind the sphere}
\label{back_view}
\end{figure}

The present research was carried out in a low-velocity water channel of the PMMH laboratory using the LIF (Laser Induced Fluorescence) visualization and the PIV (Particle Image Velocimetry) method. The cross-section of the channel was $10 \times 10 \rm{cm}$. The length of the test section was $86 \rm{cm}$. The diameter of the sphere was $D=2 \rm{cm}$. The sphere was rotating around an $x$-$axis$ parallel to the free-stream flow (see Fig. \ref{ex}). The investigated velocities were between $0.4$ and $1.8 \rm{cm/s}$, while the corresponding Reynolds number was between 75 and 350. The characteristics of the flow channel and the PIV system are also well discussed in previous papers on the wake shedding behind 3D bodies. See Klotz \textit{et al.} \cite{Klotz} and Chrust \textit{et al.} \cite{Chrust} for a full description.

The angular velocity of the sphere was controlled with high accuracy by an immersed in the fluid, brushless electric motor. As indicated earlier, the main difficulty consisted in delivering a rotational drive system, which would introduce negligible perturbation into the flow. This system was designed not only to rotate the sphere but also to inject the colorant for the visualization. To protect the motor both from flooding as well as from damage a special cover with seals was manufactured. The motor itself was placed inside the water channel at a considerable distance ($100 \rm{cm}$) upstream from the test section in order to reduce the flow perturbation. The torque was transmitted by the $0.3 \rm{cm}$ carbon fiber high-stiffness tube to reduce the effect of vibration. All parts were assembled using the laser pointer and the camera to maintain coaxiality. The investigated angular velocity remained in the range of 0-200 rpm. The misalignment was estimated to be smaller than $0.01\rm{cm}$.

To visualize the flow around the rotating device (see Fig. 1), the dye (fluorescence dye in solution) was injected into the flow through a small hole in the middle of the sphere. Additional, dedicated parts were manufactured in order to resolve the difficulties of the dye injection simultaneously with the rotation (see Fig. 1 for the detailed description). Within this system, the fluorescein passes from the external reservoir to the rotating sphere as follows: (i) by the silicone tube and the steel tubes to the small bullet shape body located on the axis of rotation, (ii) then into the rotating perforated tube, (iii) next it goes outside of the bullet shape body in the steel tube and (iv) at the end reaching the sphere where a thin slot allows the outflow of the dye. Special seals have been used to prevent leaking between the rotation axis and the fixed bullet body.

2D Particle Image Velocimetry method was used to obtain quantitative data related to the flow behind the rotating sphere. This well-known, non-intrusive, indirect measurement technique allows obtaining two in-plane velocity components of the investigated flow field. The water was seeded with tracer particles with the mean diameter of 11 $\rm{\mu}$m. The particles were carried by the fluid and their position was determined by illumination with the laser sheet and captured by a high-resolution digital camera with a possibility to measure 15 fields per second. A back view setup was used in this study as the water tunnel has a special transparent window at the end of the test section \cite{Gumowski}. For the back view measurements, the laser slice was placed behind the sphere at the distance of $x=2D=4\rm{cm}$ and oriented orthogonally to the streamwise velocity (Fig. \ref{back_view}). A standard two-dimensional image PIV set-up was used: an ImagePro $1600 \times 1200$ 12 bit CCD camera, Nikkor f70-180 mm lenses, a Minilite ND:YAG, double-pulsed laser (software and hardware were delivered by LaVision\textsuperscript{\textregistered}). For each case, 690 pairs of snapshots at 15Hz were taken, which was equivalent to 46 seconds of recording, which assures sufficient time for frequency analysis of Section \ref{sectionRD}. For all the PIV measurements, we used an interrogation window of $32\times32$ pixels with an overlap of 50\%. The field of $100\times74$ velocity vectors was obtained from each snapshot pair. The grid spacing for velocity field was $dy=dz=0.56\rm{mm}$. Subsequently, a streamwise vorticity component was calculated from the measured velocity of the flow, by numerical derivation.The measurement error of velocity was estimated to be below 5\% \cite{Klotz}.

\section{Control parameters}

The measurements focused on the identification of the flow regimes that appear depending on the values of two control parameters, namely the Reynolds number of the sphere $Re=v_{\infty}D/\nu$ and the dimensionless rotation rate or swirl parameter $\Omega=v_{\theta}/v_{\infty}$ ($D$ is the diameter of the sphere, $v_{\infty}$ stand for the upstream free stream velocity, $\nu$  denotes the kinematic viscosity, while $v_\theta$ is a maximum azimuthal velocity of the sphere). Alternatively, we consider also the Rossby number $Ro=1/{2\Omega}$, which denotes the ratio of inertial to Coriolis force.

\begin{figure}
\centerline{\includegraphics[width=18 pc]{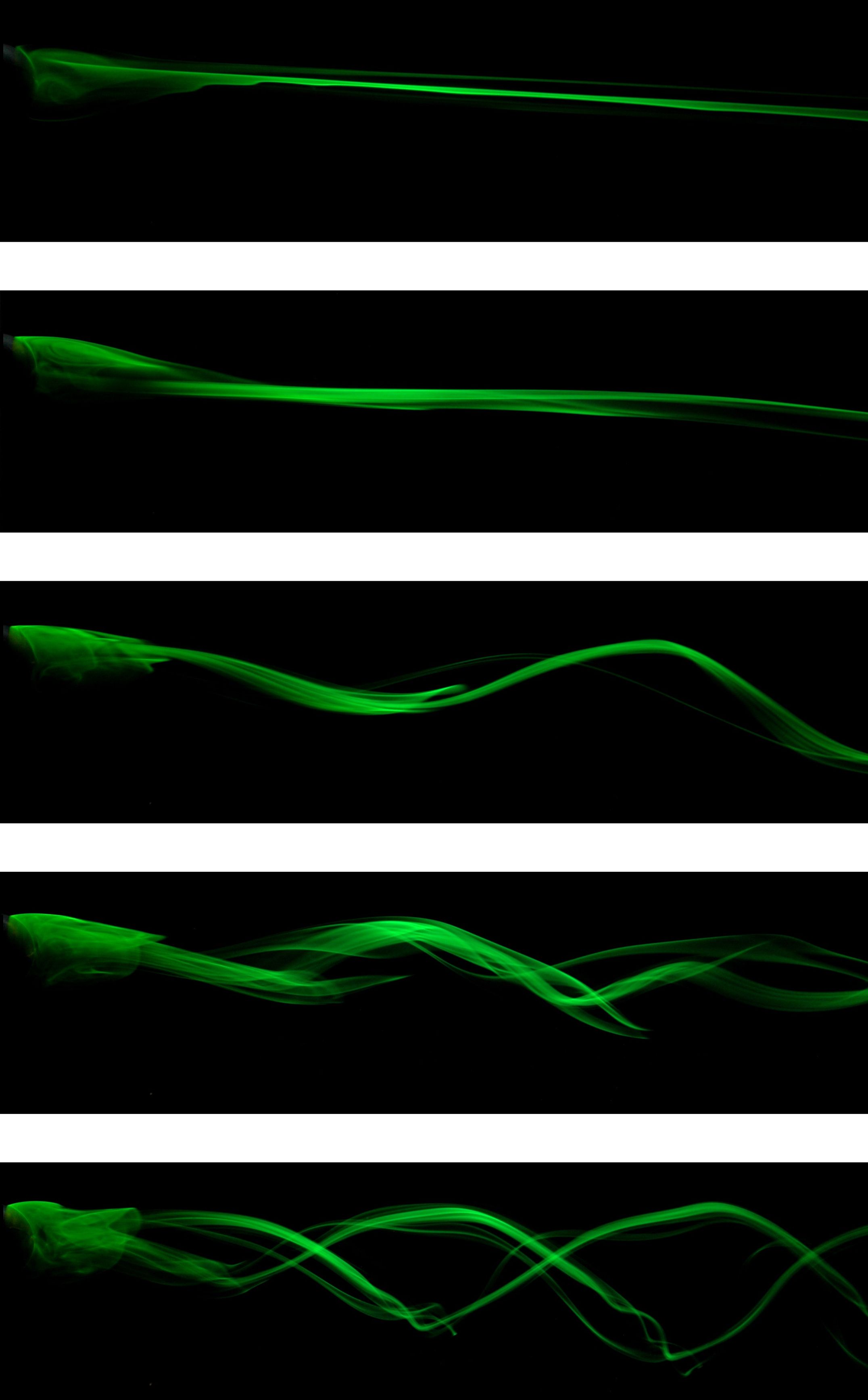}}
\caption{Side view visualisation for $Re = 250$. From top to bottom: $\Omega$ = 0 (\textit{2CRV} regime), 0.2, 0.5 (\textit{low helical}), 0.8, 1.2 (\textit{high helical}).}
\label{vis250}
\end{figure}

\section{Flow scenarios - qualitative study}
\label{sectionVis}

\begin{figure}
\centerline{\includegraphics[width=18 pc]{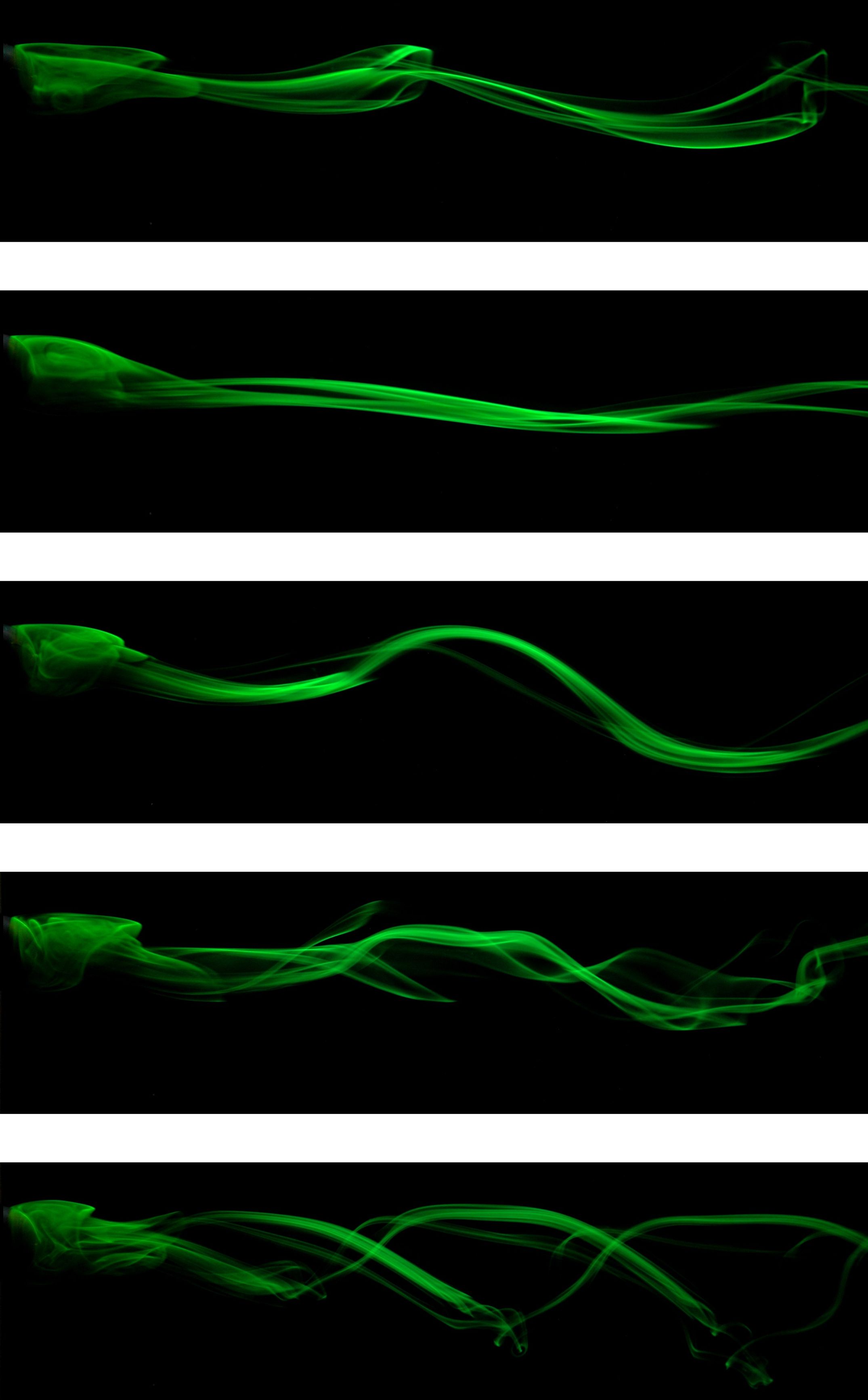}}
\caption{Side view visualisation for $Re = 300$. From top to bottom: $\Omega$ = 0 (\textit{HS} regime), 0.2, 0.5 (\textit{low helical}), 0.8, 1.2 (\textit{high helical}).}
\label{vis300}
\end{figure}

To demonstrate the possible flow scenarios, the side visualizations are presented in Figs. \ref{vis250} and \ref{vis300}. In this part we have focused on the steady and the unsteady cases which represent two different regimes for the static ($\Omega$ = 0) sphere: the steady planar symmetric flow with two weak counter rotating vortices (\textit{2CRV}) and the periodic hairpin vortex shedding (\textit{HS}). It was expected that the observed wake for $\Omega>0$ will depend on the initial regime. This appeared to be the case, for the first, axisymmetric flow regime (i.e., $Re=100$), for which the wake remains axisymmetric for rotation rates $0 < \Omega < 2$. On the other hand, for Reynolds numbers greater than $350$ and for rotation rate greater than 4 the wake becomes chaotic or very complex. Therefore the natural limitation of our research is $Re<350$ and $\Omega<4$.

Explored Reynolds numbers $250$ and $300$ correspond to two different initial regimes (\textit{2CRV} and \textit{HS}). We present here the most representative visualization results: $Re=250$ and $Re=300$ and $\Omega$ = 0, 0.2, 0.5, 0.8, 1.2. Visualization patterns are very similar for $Re=250$ and $Re=300$ and for corresponding $\Omega$ (the difference is observed only for $\Omega=0$ which represents the state of \textit{2CRV} for $Re=250$). Nevertheless, it is important to notice that this is only a visualization (representing the streaklines in an unsteady process) and the far-reaching conclusions should not be drawn at this point. A typical pattern of \textit{HS} (hairpin shedding) is depicted in the Fig. \ref{vis300} for $\Omega=0$ and $Re = 300$. 

For $\Omega = 0.2$ we observe one large vortex while the second, weaker one, is slightly waving. With the growing rotation rate ($\Omega$ = 0.5) a different behavior of the wake can be observed, a single vortex apparently spirals around the axis. This state we call \textit{low helical}. Next, for $\Omega$ = 0.8 (and $Re = 250$) the wake becomes unsteady, while the observed pattern becomes irregular. Beyond this state, for $\Omega = 1.2$, the flow becomes regular once more and becomes similar to the helical state but with two helical branches. We call these similar states \textit{low helical} (\textit{LH}) or mode $m=1$ in the case of $\Omega=0.5$ and \textit{high helical} (\textit{HH}) or mode $m=2$ for $\Omega=1.2$. 

Similar visualizations were obtained also for the initial \textit{HS} regime at lower Reynolds number ($Re = 300$). We observe differences between $Re = 250$ and $Re = 300$ in the wakes for $\Omega$ = 0, however, these differences disappear for the larger $\Omega$ namely 0.2, 0.5, 0.8 and 1.2. For $\Omega = 0.2$ and $Re = 300$ the hairpin shedding instability is no longer observed. These flow phenomena and bifurcations are investigated in detail and described in further Sections.

\begin{figure}
\centerline{\includegraphics[width=22 pc]{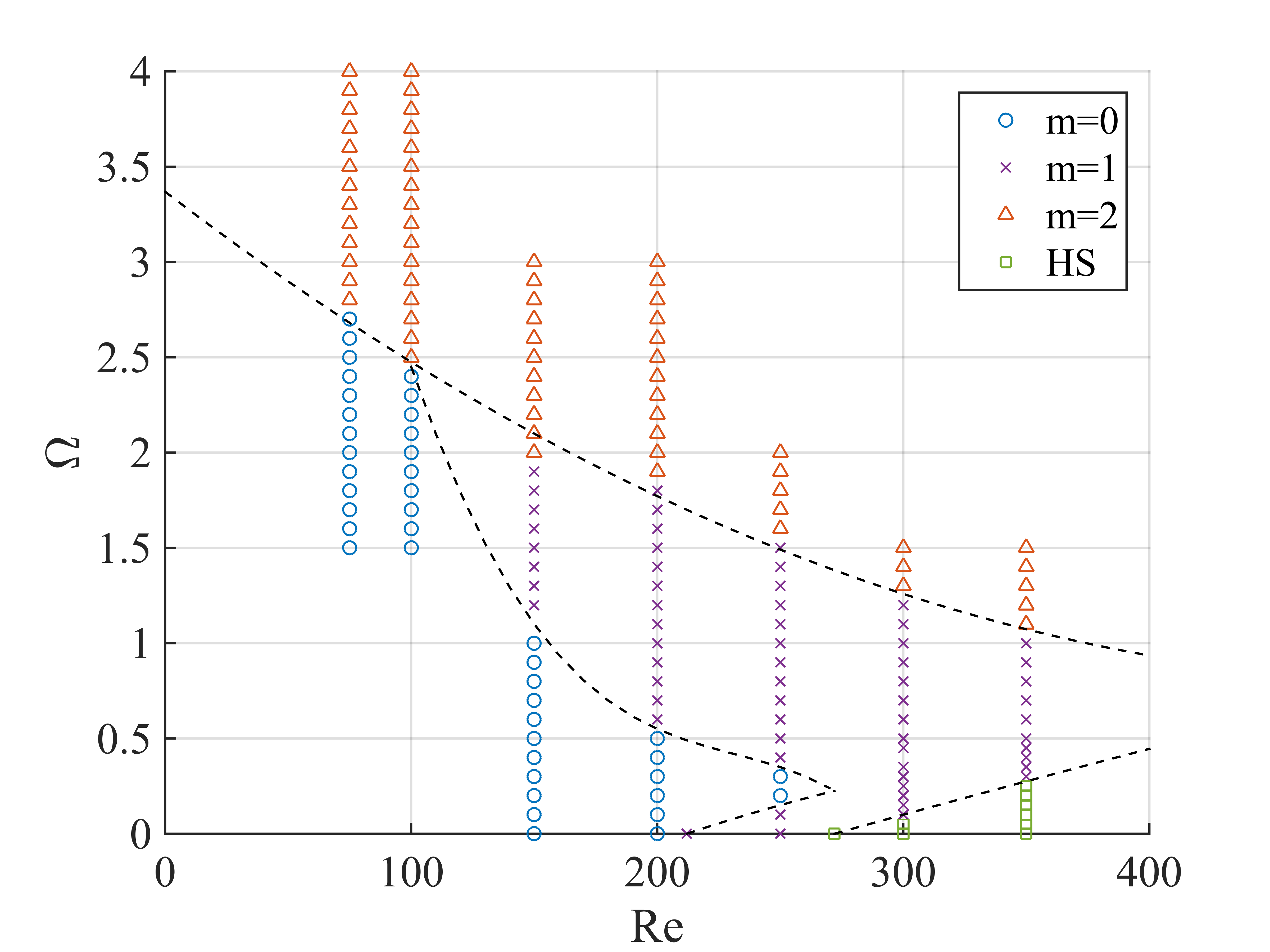}}
\caption{Full bifurcation map in the $(Re,\Omega)$ - parameter plane. The boundary between regions has been estimated by modal analysis and frequency analysis of the PIV measurements. The different predominant modes correspond as follows: $m=0$ axisymmetric flow, $m=1$ \textit{low helical}, $m=2$ \textit{high helical} and \textit{HS} stands for hairpin shedding instability. Each point represents a single PIV measurement.}
\label{map}
\end{figure}

We present in Figs. \ref{vis250} and \ref{vis300} an overview, by visualization, of the different flow regimes. In addition, Fig. \ref{map} shows the parametric map of the existing flow regimes. This map is obtained not from the visualizations but by using different quantitative methods detailed in the next Sections. Note that minor differences exists between visualization results and PIV results, for instance, the \textit{HH} state in visualization was seen for $Re=250$ and $\Omega=1.2$ whereas PIV, modal and frequencies analysis detected the bifurcation point at $\Omega=1.5$.

\section{Quantitative analysis of results}
\label{PIVres}

\begin{figure}

\subfigure[]{
\includegraphics[width=35 pc]{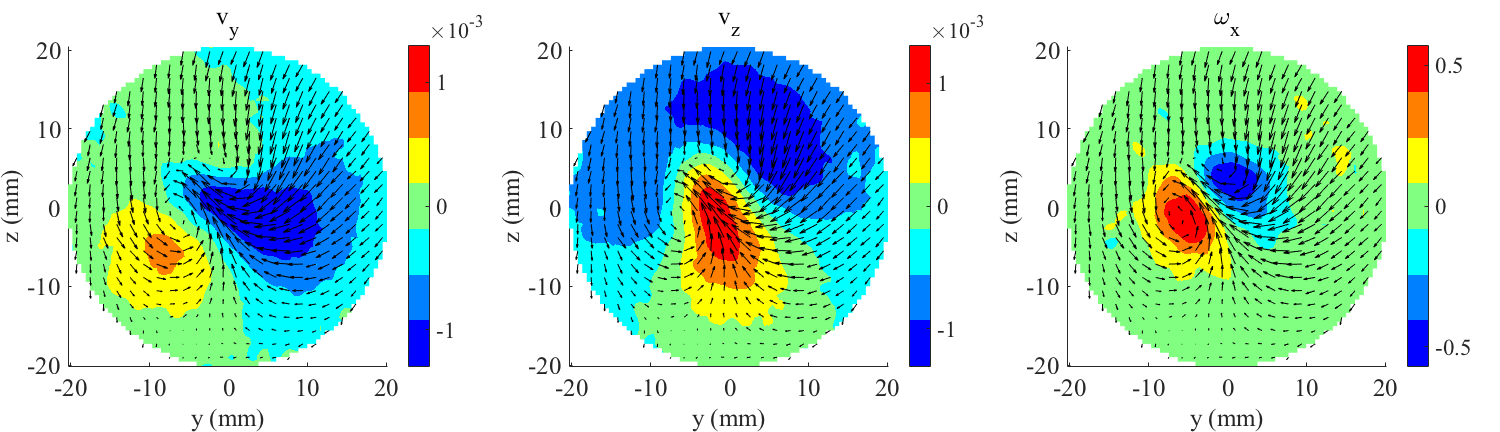}
}
\subfigure[]{
\includegraphics[width=35 pc]{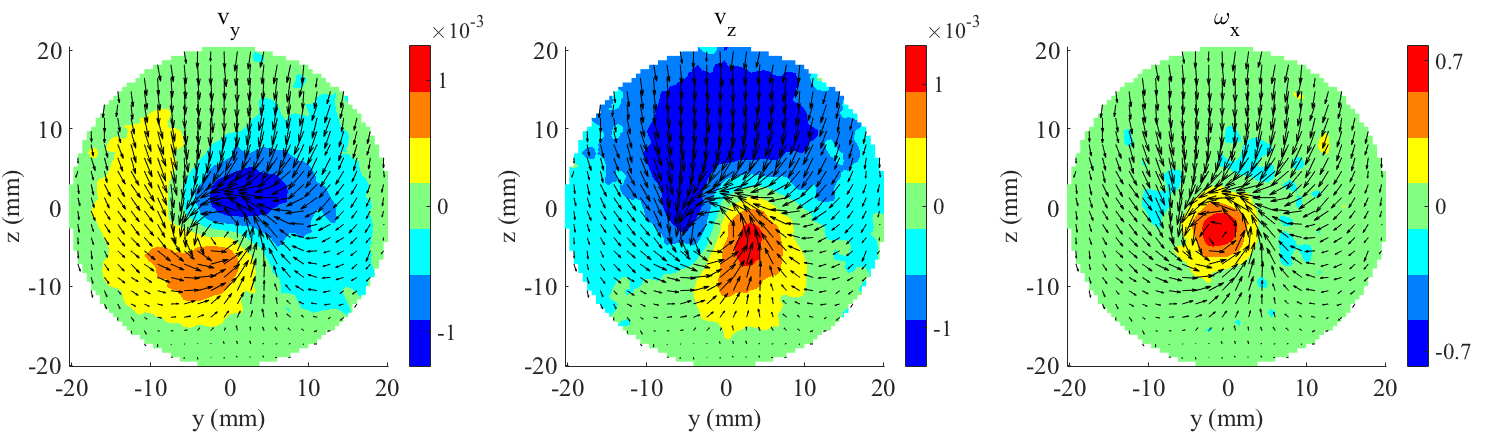}
}
\subfigure[]{
\includegraphics[width=35 pc]{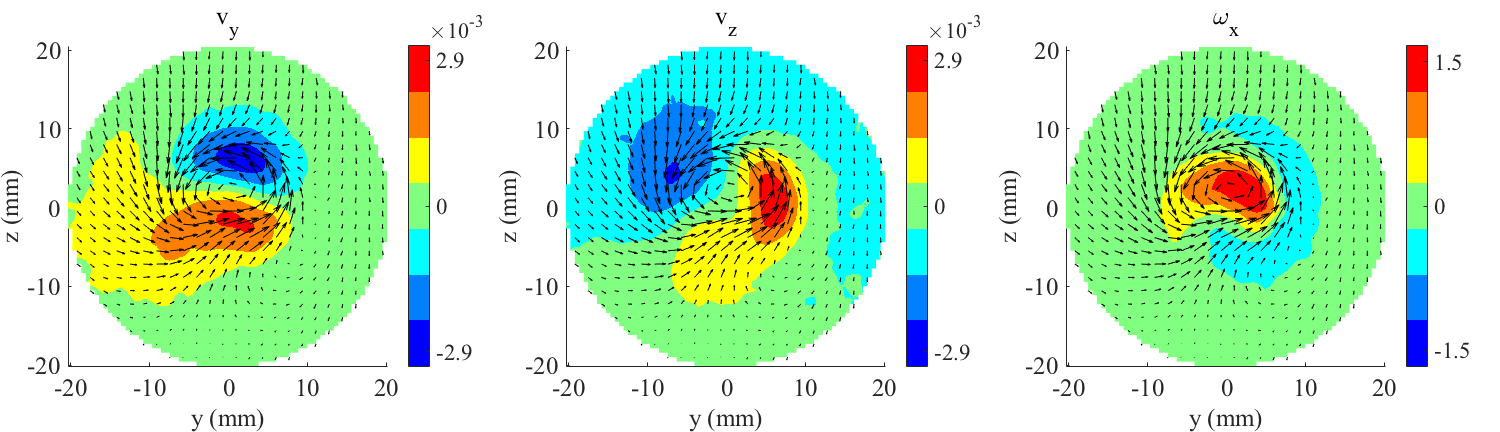}
}
\subfigure[]{
\includegraphics[width=35 pc]{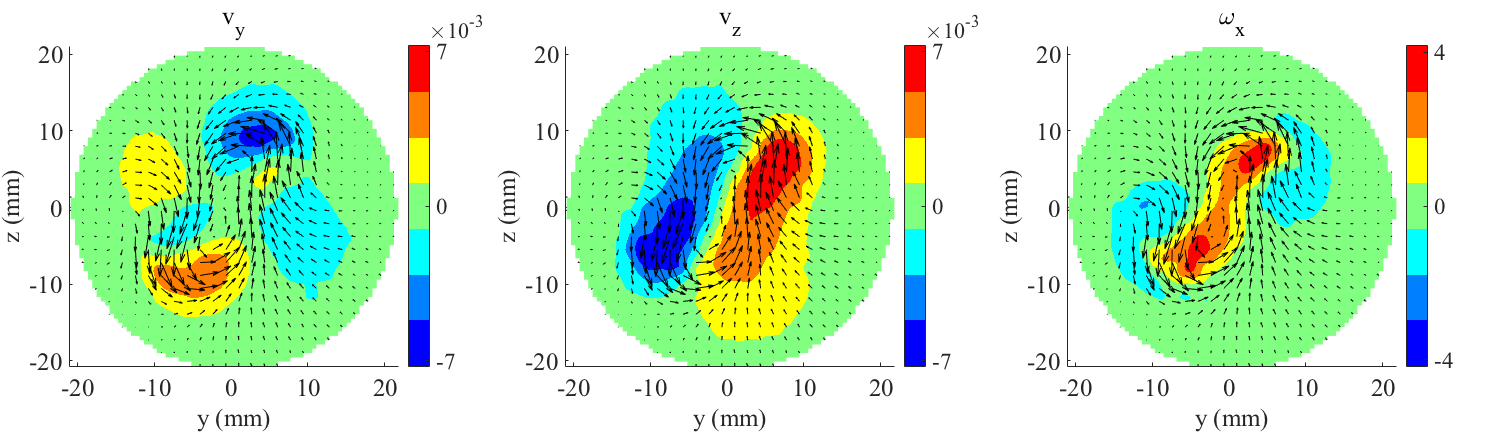}
}
\caption{Instantaneous velocity and vorticity fields from PIV measurements for $Re=250$. In each row the  background color represents velocity components $v_y$ (m/s), $v_z$ (m/s), and the longitudinal vorticity $\omega_x$ (1/s) (from left to right). The consecutive rows correspond to a different rotation rate: (a) $\Omega=0$, \textit{2CRV}, (b) $\Omega=0.2$, axisymmetric, (c) $\Omega=0.5$, \textit{LH}, (d) $\Omega=1.6$, \textit{HH}}
\label{velocity_more}
\end{figure}

\begin{figure}
\centerline{\includegraphics[width=15 pc]{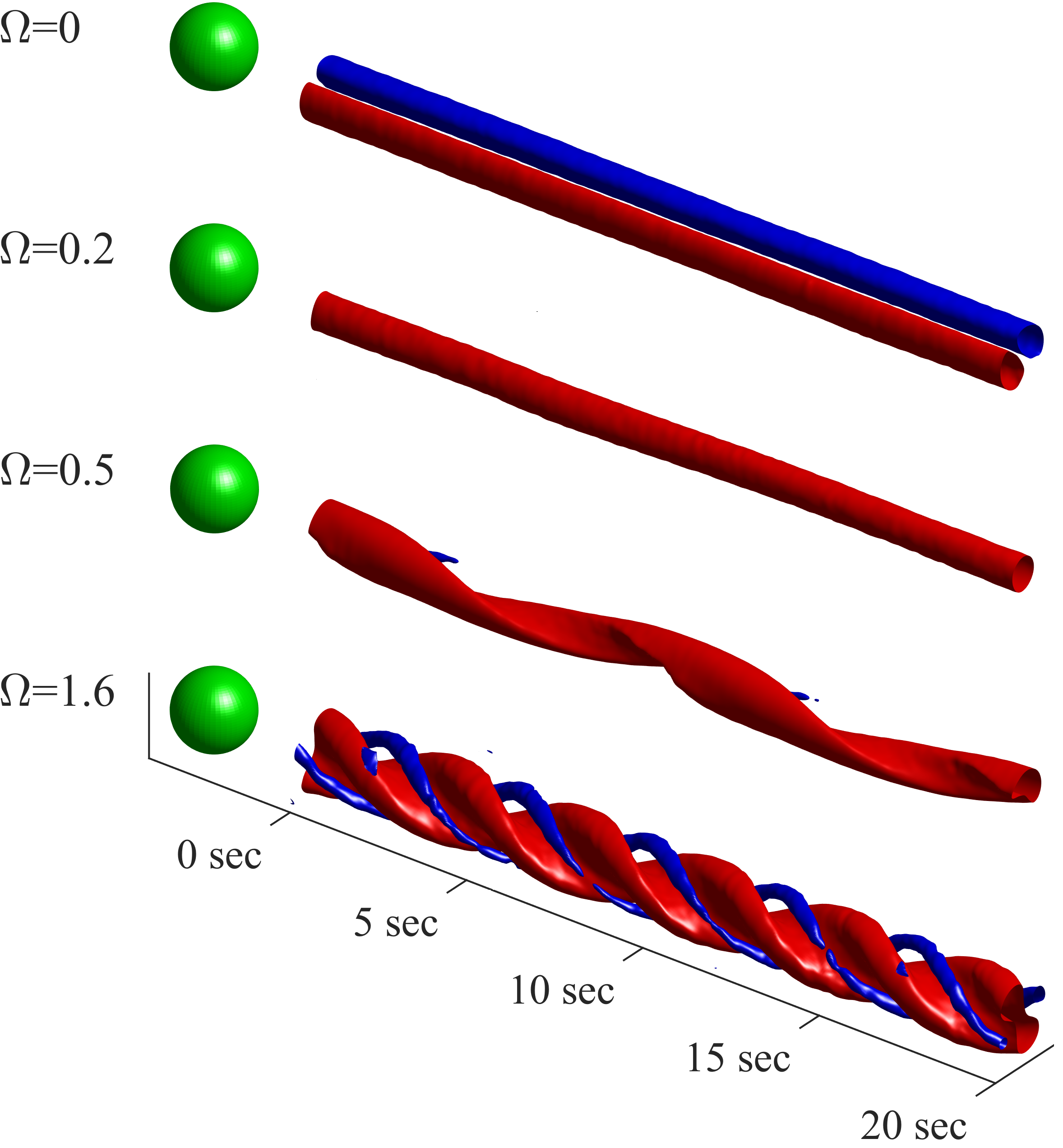}}
\caption{Spatio-temporal representation of the measured streamwise vorticity at $x=2D$ behind the sphere. $Re=250$, $\Omega=0$ (\textit{2CRV}), $0.2$ (axisymmetric), $0.5$ (\textit{LH}), $1.6$ (\textit{HH}) from top to bottom. The red and blue isosurfaces correspond to $\omega_x=\pm0.3\omega_{max}$ respectively, where $\omega_{max}$ denotes the maximum value of vorticity for each case. Horizontal axis corresponds to the elapsed time.}
\label{piv}
\end{figure}

The quantitative results are obtained by PIV measurements (see example in Fig. \ref{velocity_more}), but it is worth to notice, that because only two (transversal) components of velocity are measured, only single component of vorticity - namely its longitudinal or streamwise component $\omega_x$  - can be calculated. Nevertheless, this information is sufficient to identify the relevant flow phenomena in the wake \cite{Klotz,Szaltys}. Because for almost all investigated cases the flow is time dependent, we present results using time-space reconstruction. For the purpose of analysis, the longitudinal component of vorticity $\omega_x(y,z,t)$ was measured at $x=2D$ behind the sphere downstream of the flow, already outside the recirculation bubble ($y,z$ are the coordinates orthogonal to the free stream velocity while $t$ denotes time). In order to illustrate the evolution of vortical structures ``in time'' at a given transversal plane $x=2D$ rather than ``in space'', we present in Fig. \ref{piv} the isovalues of $\omega_x$ colored by its sign (red positive, blue negative)

This set of four time-space reconstructions is based on the PIV measurements. The most representative examples were chosen to show how the appearing regimes depend on the rotation rate $\Omega$. The results are presented for $Re=250$ and $\Omega=0, 0.2, 0.5, 1.6$. Four, clearly different, consecutive states are presented: (i) the two steady, equal, counter rotating vortices (\textit{2CRV}), (ii) the axisymmetric flow (mode $m=0$), (iii) the \textit{low helical} (\textit{LH})or spiral mode (mode $m=1$) and (iv) the \textit{high helical} (\textit{LH}) or mode $m=2$. More information about the evolution of the flow field can be inferred by the subtraction of the pure rotation (mode $m=0$ of the Fourier decomposition).

For further analysis, we separate the flow field into parts representing different phenomena. For the static case $\Omega=0$ the flow is planar symmetric for the steady ($Re>Re_1$) as well as for the periodic regimes ($Re>Re_2$). On the other hand for $\Omega\ne 0$ the rotation apparently induces the cyclonic vorticity in the wake (this is confirmed both by visualizations as well as  PIV measurements). These two observations suggest that the wake can be considered as sum of two or more components. These components we can identify using the  azimuthal Fourier decomposition. The result is analyzed as a function of Reynolds number and $\Omega$. 

Originally the longitudinal vorticity field $\omega_x(y,z,t)$ is available from the PIV measurements on a Cartesian grid $(y,z)\in\mathbb{R}^2$ for a fixed plane $x=const$. This representation is interpolated to the polar grid $(r,\theta)\in\mathbb{R}^2$. Subsequently a sequence of 1D polar Fourier transforms are calculated, (see \citet{Szaltys} for details):
$$\lambda(r_k,m)=\sum_n \omega_x(r_k,\theta_n)e^{im\theta_n}$$
$$\theta_n = n\frac{2\pi}{N},\, n\in(0...N-1)$$
$$r_k = k\frac{r_{ext}}{K},\, k\in(1...K)$$

\noindent where $r_{ext}$ denotes the external radius of data used in this analysis ($r_{ext}=1.5\rm{cm}$ in our case) as for $r>r_{ext}$ the vorticity is negligible.

Finally, the amplitudes of azimuthal modes are obtained by numerical integration in the radial direction.
$$\lambda(m)=\frac{1}{\sum_k r_k\Delta r}\sum_k \lambda(r_k,m)r_k\Delta r$$

\begin{figure}
\centerline{\includegraphics[width=18 pc]{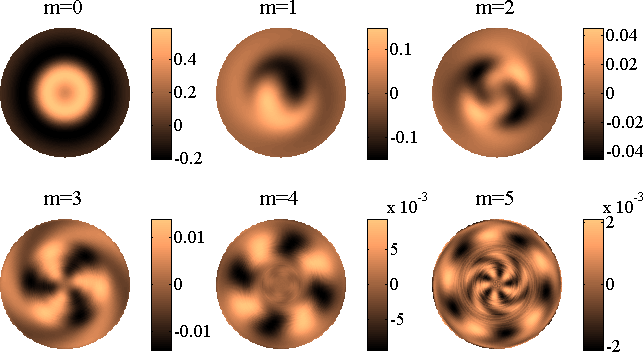}}
\caption{Fourier Azimuthal Decomposition patterns for instantaneous field of $Re=250$ and $\Omega=0.5$ (each mode has its own colour scale)}
\label{modes}
\end{figure}

The main mode patterns (from $m=0$ to $m=5$) obtained by the Fourier decomposition of the instantaneous vorticity field are presented in Fig. \ref{modes} for $Re=250$ and $\Omega=0.5$. The first 6 modes represent about 99.9\% energy of the entire vorticity field, while modes $m=0,1,2$ contain 99.4\% of all energy. As it will become evident, these 3 modes are sufficient to describe the behavior of the wake, as the evolution of these modes follows the evolution of flow. Further analysis is therefore focused on these first 3 modes (separately for each pair of control parameters $(Re,\Omega)$). It is worth to note that the spatial decomposition is performed for each instantaneous flow field and as a consequence, we obtain the decomposition for each time step.

Table 1 shows the time-space reconstruction, for $Re=250$ and $\Omega=0.5$ and $1.6$, for the filtered $\omega_x$ consisting of modes $m=1,2$ as compared with the full vorticity field, $\omega_x$. This improves the clarity of presentation as the axisymmetric mode $m=0$, which represents the rotation of the whole system, in this way remains hidden. 

For $\Omega>0$ we have either the axisymmetric flow or the rotating vortical patterns, therefore the flow average in time consists mostly of the mode $m=0$. The higher mode $m=1$, appears primarily for the static case  ($\Omega=0$) and for $Re>Re_1=212$. The decomposition into azimuthal modes is very sensitive to the selection of the zero point in the polar coordinates system and may produce unsatisfactory results if this point is improperly selected (in particular the sphere $x$-axis is not good enough for this purpose). Therefore, in the present paper, we have selected the position of the zero point by the optimization process, in which the number of modes representing the flow field was minimized. In most cases, we have achieved no less than 98\% contribution of mode $m=0$ in the averaged field.

\begin{table}[t]
\caption{Spatial-temporal representation of the filtered streamwise vorticity. We present the sum of the modes $m=1$ and $m=2$ (right) as well as a comparison with the full vorticity field (left). $Re=250$, $\Omega=0.5$ (top), $\Omega=1.6$ (bottom). (The red and blue isosurface corresponds to $\omega_x=\pm0.3\omega_{max}$ respectively, where $\omega_{max}$ denotes the maximum value of vorticity for a given case)}
\label{tabela}
\centering
 \begin{tabular}{M{16 pc}|M{16 pc}}
\hline
 & \\
Full vorticity field & Filtered vorticity field (without $m=0$) \\

\includegraphics[height=7 pc]{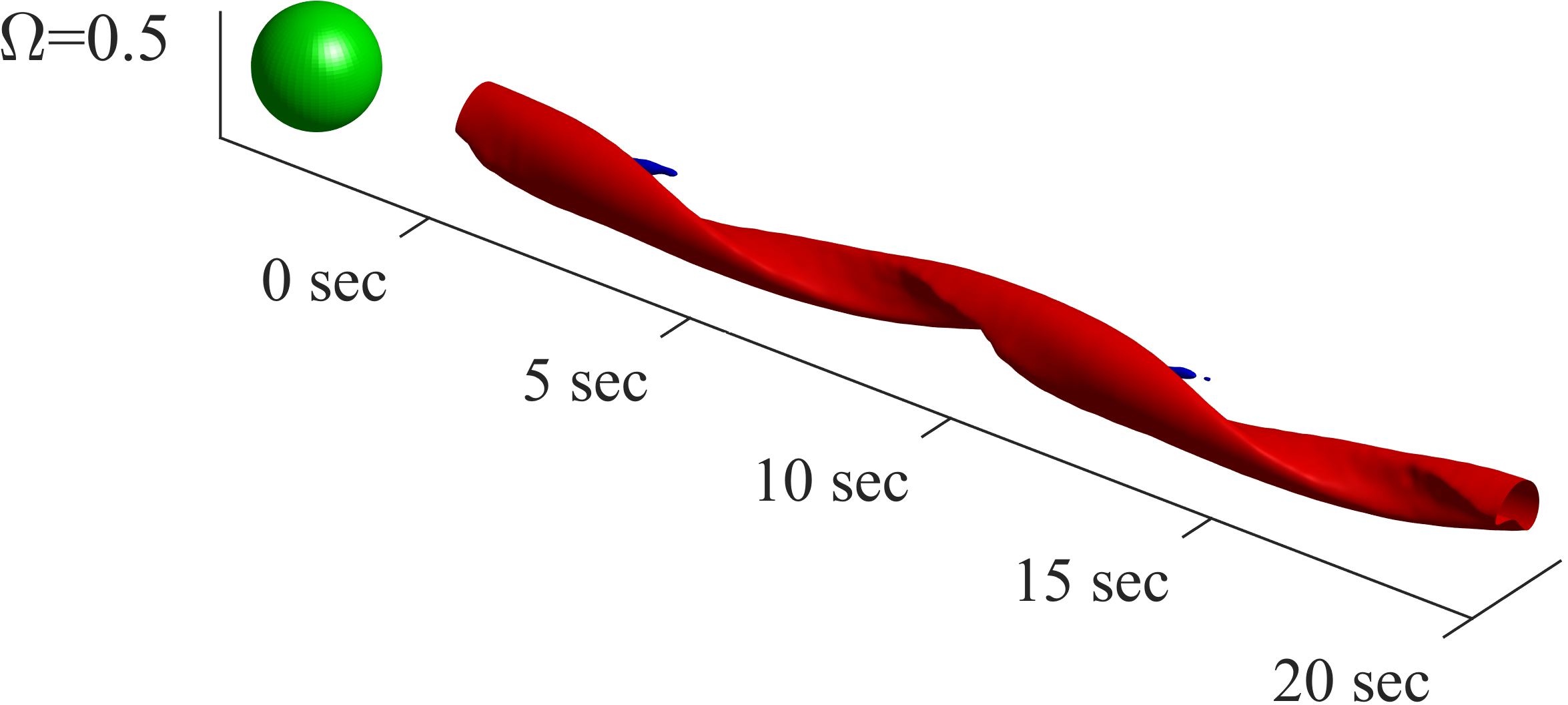} & \includegraphics[height=7 pc]{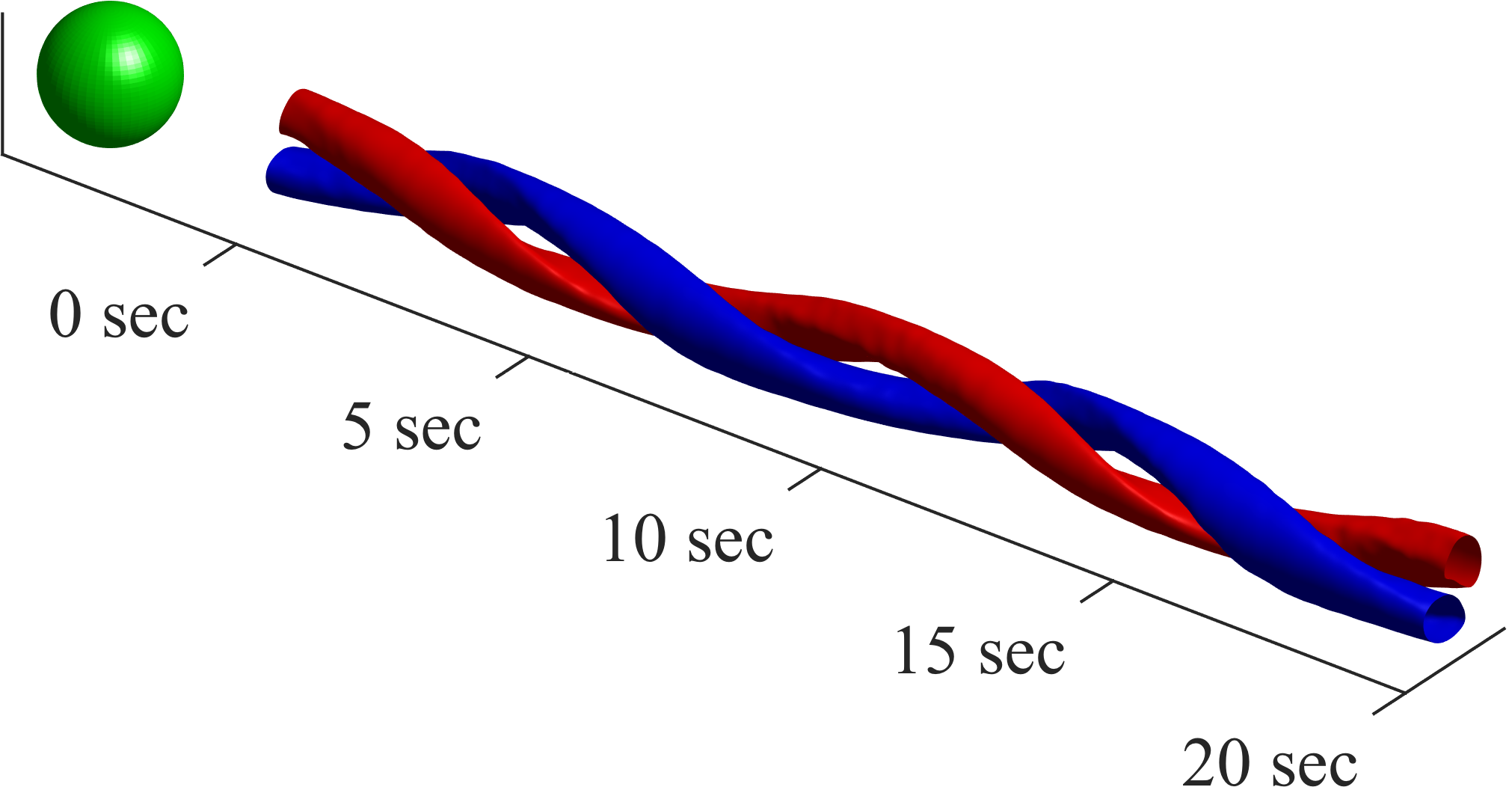}\\

\includegraphics[height=7 pc]{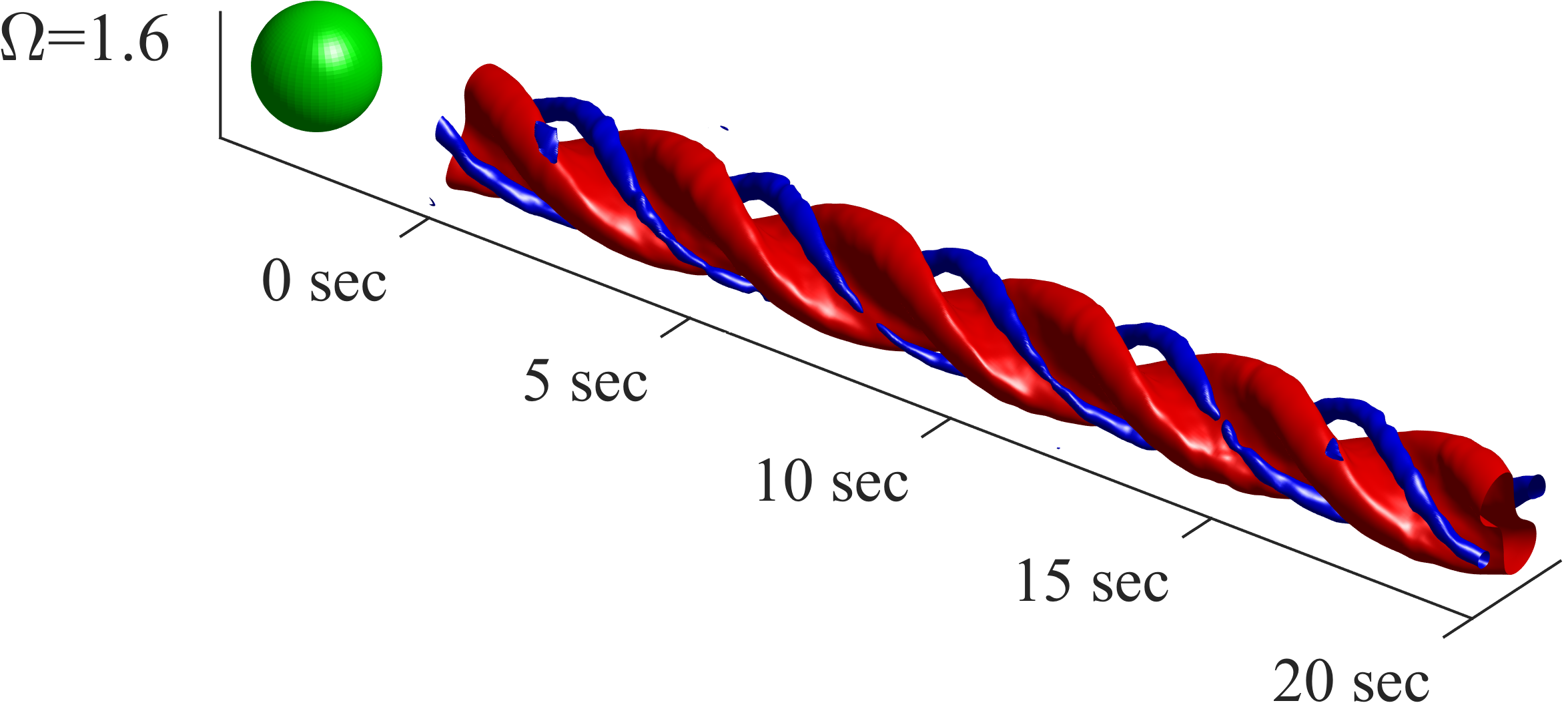} & \includegraphics[height=7 pc]{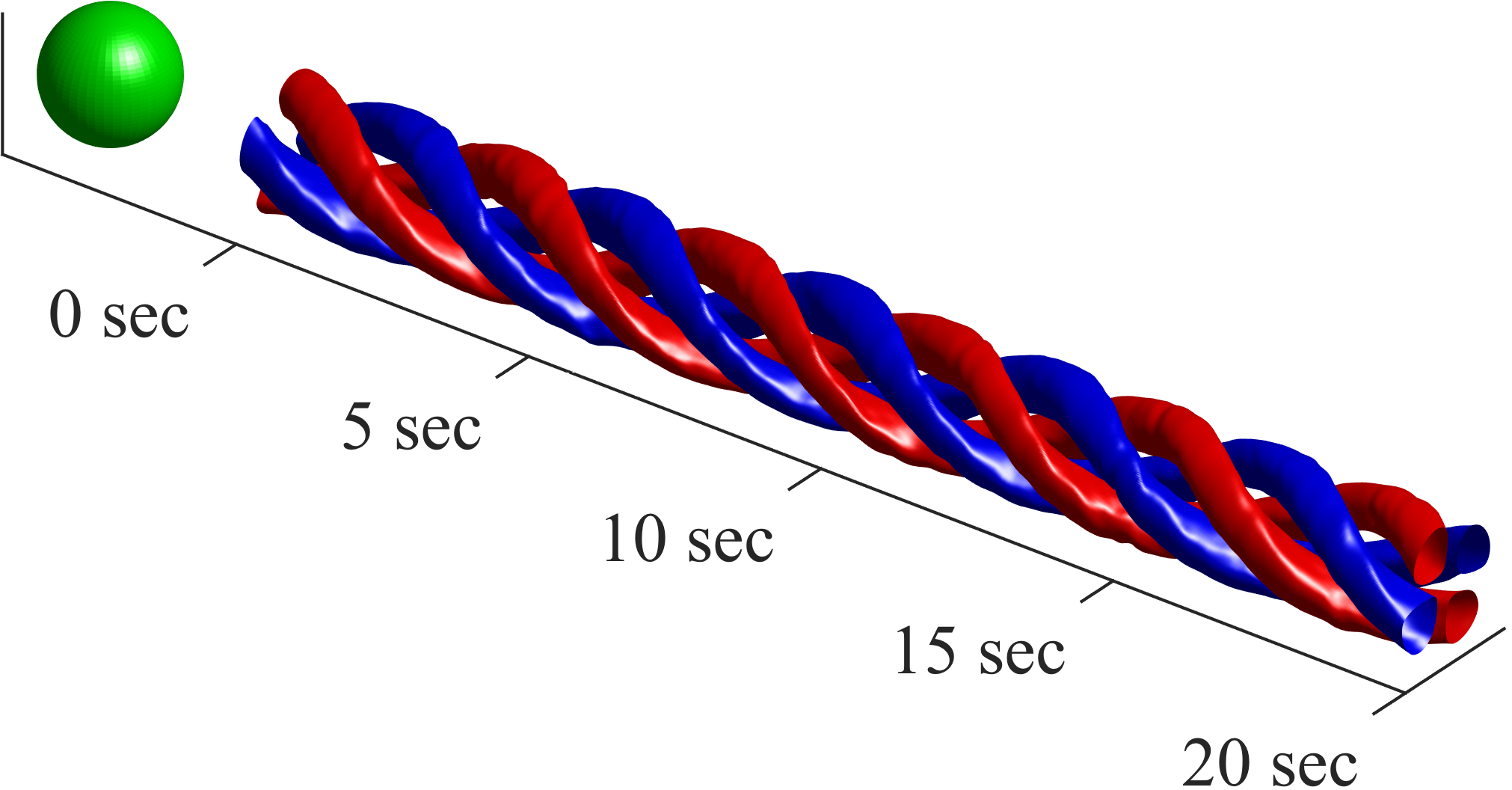}\\
  \hline
\end{tabular} 

\end{table}

The investigated flow remains nearly periodic either because of the vortex shedding or because of the rotation of the sphere (this could also include the possible off-design effects of axis misalignment). All our previous observations indicate that the corresponding characteristic frequencies will appear in the wake. To identify these frequencies (and their evolution) we use the Fast Fourier Transform (FFT) and finally the Dynamic Mode Decomposition (DMD).

This FFT analysis is applied in the time domain to the streamwise vorticity $\omega_x(x,y,z,t)$ at the cross-section $x=2D$ and at a particular point $y_{\star}, z_{\star}$.  In order to filter out the noise, the result is averaged over all points lying within the $d_\star$ around $(y_\star,z_\star)$. To achieve as monochromatic spectrum as possible the values of $y_\star, z_\star,d_\star$ are selected in a process of optimization. The triple ($y_\star, z_\star,d_\star$) is different for each pair $(Re,\Omega)$ but in majority of cases (more than 75\%) we obtain $d_{\star}=2.8$--$3.36\rm{mm}$ which corresponds to $25-36$ points of discrete PIV grid.

This approach leads to identification of dominant frequencies in the wake and reduction of noise and harmonics. The obtained spectra (Figs. \ref{f75-100}, \ref{f150-200}, \ref{f250} and \ref{f300-350}) are represented as colorbars for which red denotes the highest amplitudes, while white stands for the zero amplitude. The maps of spectra consisting of different values of $\Omega$ are generated for different values of $Re=$ 75, 100, 150, 200, 250, 300, and 350. In all figures the frequency $f$ is presented as a dimensionless Strouhal number $St=fD/v_{\infty}$. The dashed line represents the rotational frequency of the sphere. The residual existence of this frequency in the spectrum may be caused by the possible misalignment of the rotation axis.

\section{Results and discussion}
\label{sectionRD}
The purpose of this part of analysis is to study in detail the evolution of modes and frequencies as a function of $\Omega$ and the Reynolds number. We present the streamwise enstrophy of each mode, as the order parameter of the flow evolution. The streamwise enstrophy is defined as $\epsilon=\frac{1}{|S|}\int\omega_x^2dS$ and measures the energy of the flow field carried by a particular mode ($S$ denotes the area over which the vorticity is integrated - in our case it was the circle $r=1.5\rm{cm}$). One should note that by the ``mode'' we understand here an artificial component of the instantaneous vorticity field, which, due to the nonlinearity of the Navier-Stokes equations, may not have a physical representation of a separate flow feature (in contrast to the normal modes). Nevertheless, the sum of these artificial components gives full information about the original field. In Figs. \ref{Re75a}, \ref{Re100a}, \ref{allM0}, \ref{Re150a}, \ref{Re250a}, \ref{Re300a}, and \ref{Re350a} the square-root of enstrophy of the mode $m=0$ is presented as a function of $\Omega$. This quantity was further nondimensionalised by multiplication by $D/v_\infty$. The square root was selected in order to assess the proportionality to the vorticity $\omega_x$, and therefore for comparison with the frequency of the sphere's rotation ($f_{rot}=\frac{v_\infty}{D}\frac{\Omega}{\pi}$ or $St_{rot}=\frac{\Omega}{\pi}$). The linear proportionality (of $m=0$) was observed in the range of flow velocities corresponding to stationary or periodic flow properties.
In the following the relation of different flow patterns and the behaviour of modes is discussed in detail.

The discussion on behaviour of modes and frequencies is organized by the existing flow regimes, namely, axisymmetric flow, \textit{low helical}, \textit{high helical}, and hairpin vortex shedding. For the purpose of clarity of the presentation the main notations are recalled here: $Re_1$ and $Re_2$ stand for first two bifurcations of static sphere, 212 and 272 respectively. The characteristic frequency observed in the spectrum is different for each regime (\textit{LH, HH, HS}) and depends on $Re$, therefore for Strouhal numbers we use following notation: $St^1_{Re}$, $St^2_{Re}$, and $St^{HS}$, where upper index, (1), (2), and (HS), stands for regime, i.e., \textit{low helical}, \textit{high helical}, and hairpin shedding, respectively, while lower index denotes Reynolds number.

\subsection{The axisymmetric flow and mode $m=0$}

\begin{figure}
\hfill
\subfigure[mode m=0 of the average field]{
\label{Re75a}
\includegraphics[width=\plotsw pc]{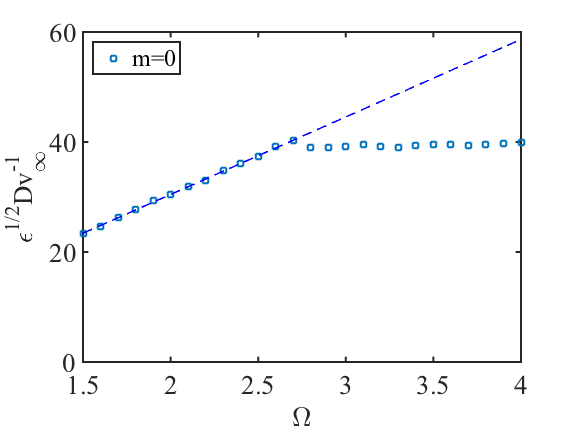}
}
\hfill
\subfigure[modes m=1,2,3,4]{
\includegraphics[width=\plotsw pc]{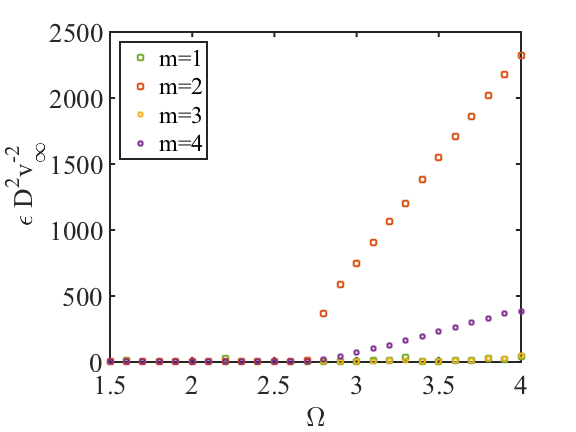}
\label{Re75b}
}
\caption{Dimensionless enstrophy of modes as a function of $\Omega$, $Re=75$. (a) mode $m=0$ grows linearly up to $\Omega=2.7$, axisymmetric flow, (b) at 2.7 modes $m=2$ and $m=4$ appear along with \textit{HH}}
\label{Re75}
\end{figure}

\begin{figure}

\hfill
\subfigure[mode m=0 of the average field]{
\label{Re100a}
\includegraphics[width=\plotsw pc]{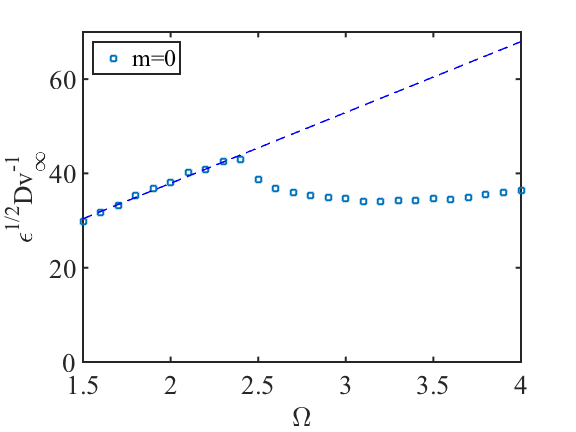}
}
\hfill
\subfigure[modes m=1,2,3,4]{
\includegraphics[width=\plotsw pc]{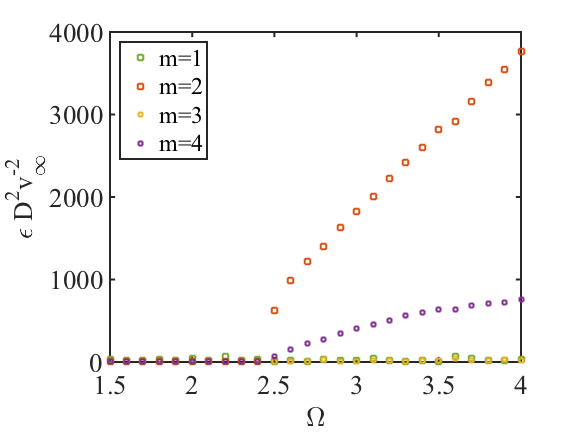}
}
\caption{Dimensionless enstrophy of modes as a function of $\Omega$, $Re=100$. The axisymmetric to \textit{HH} transition at $\Omega=2.4$}
\label{Re100}
\end{figure}

\begin{figure}
\hfill
\subfigure[spectra for $Re=75$]{
\includegraphics[width=\plotsw pc]{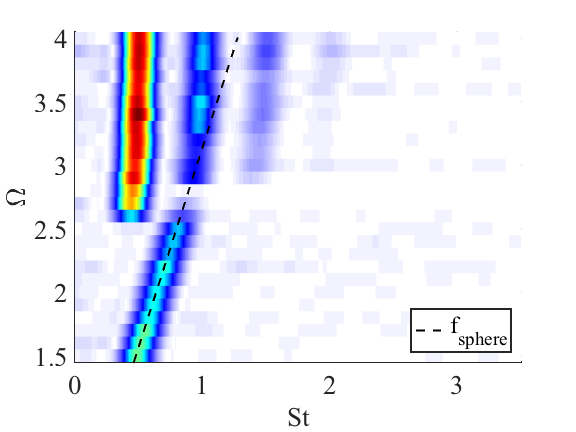}
\label{f75}
}
\hfill
\subfigure[spectra for $Re=100$]{
\includegraphics[width=\plotsw pc]{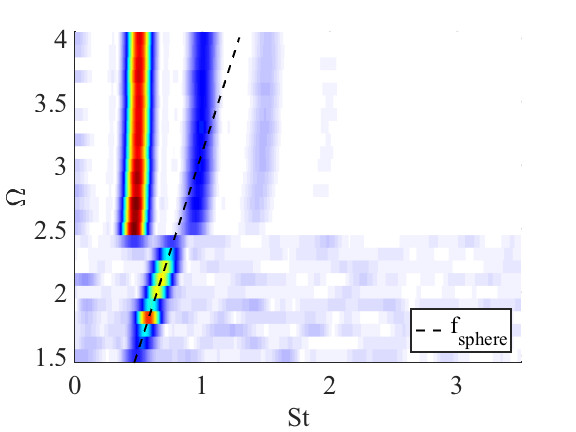}
\label{f100}
}
\caption{Spectra for $Re=75$ and $100$. The sphere rotation frequency (dashed line) was detected for axisymmetric flow up to value of $\Omega$ (a) 2.6, (b) 2.4. For higher $\Omega$ the frequency at $St^2_{75}\approx St^2_{100}=0.48$ was observed (independent from the rotation). Red and white denote the highest and zero amplitude, respectively}
\label{f75-100}
\end{figure}

For static, non-rotating sphere the axisymmetric flow exists up to critical Reynolds number $Re_1=212$. As an effect of rotation the axisymmetric flow can be destabilized resulting in the \textit{high helical} regime for low $Re$ or in the \textit{low helical} for moderate $Re$. Although axisymmetric flow does not exist for $Re>Re_1$ when $\Omega=0$, it reappears for $Re=250$ when a small value of rotation is introduced - this will be discussed in detail further on. The region of $(Re,\Omega)$ - parameter plane, for which axisymmetric flow was observed, was presented already in the Fig. \ref{map}.

The present analysis starts with $Re=75$ and $Re=100$ (see Fig. \ref{Re75} and \ref{Re100}), for which we observe a linear growth of $\epsilon^{1/2}$ for the mode $m=0$ up to $\Omega=2.7$ ($2.4$ in case of $Re=100$) and negligibly small contribution of other modes. This confirms our hypothesis that the sphere's rotation adds linearly the cyclonic vorticity to the flow field. For both cases of $Re=75$ and 100 a weak signal in the spectrum equal to the frequency of the rotation of the sphere was detected in the range of axisymmetric flow, i.e., up to $\Omega=2.6$ (Fig. \ref{f75}) and $\Omega=2.3$ (Fig. \ref{f100}).

In these cases, we have observed the influence of the forcing (rotation of the sphere) which corresponds to a non-modal (convective) character of the flow. For the other, higher Reynolds, values this character is lost in favour of clear modal or global behavior with clearly observed frequencies.

\begin{figure}
\hfill
\subfigure[mode m=0 of the average field]{
\label{Re250a}
\includegraphics[width=\plotsw pc]{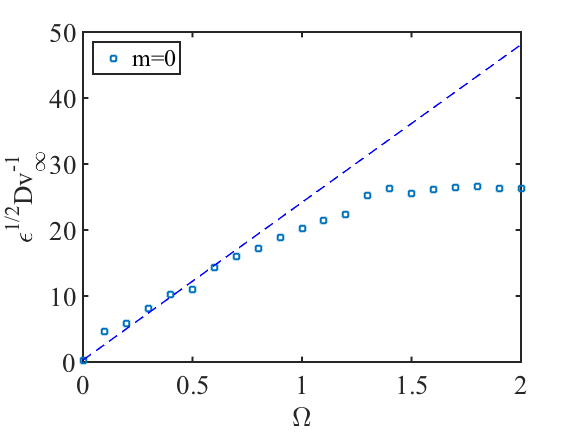}
}
\hfill
\subfigure[modes m=1,2,3,4]{
\includegraphics[width=\plotsw pc]{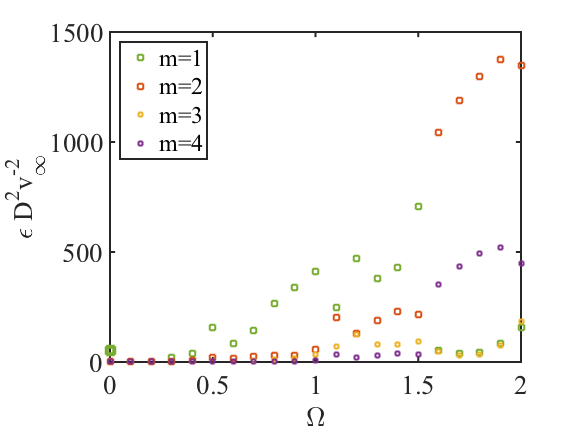}
}
\caption{Dimensionless enstrophy of modes as a function of $\Omega$, $Re=250$. (a) $m=0$ grows linearly up to $\Omega=0.6$ and stagnates at $\Omega=1.5$ (\textit{HH}), (b) an isolated point of $m=1$ of \textit{2CRV} is seen at $\Omega=0$, decays immediately as $\Omega$ increases and reappears at 0.5 (\textit{LH})}
\label{Re250}
\end{figure}

An axisymmetric flow exists also for higher $Re>Re_1$ for which initial regime was planar symmetric \textit{2CRV}. In this \textit{2CRV} regime, the vorticity remains planar-symmetric relative to the XZ plane. With increasing $\Omega$ the positive longitudinal vortex becomes stronger while the negative one weakens, as was observed already in the numerical experiment by \citet{KimChoi}. Moreover, the centre of the positive vortex moves towards the symmetry plane (as observed for $Re=250$). This effect is induced by the axisymmetric mode $m = 0$, which is added to the flow field by the fact of rotation. Eventually, the separate negative longitudinal vortex disappears, while the positive one reaches the symmetry plane and the resulting flow becomes axisymmetric. For $Re=250$, this axisymmetric regime lasts only in the narrow interval of $0.2<\Omega<0.3$, as subsequently, the positive longitudinal vortex starts waving in the $x$ direction and thus the flow becomes unsteady (see Fig. \ref{piv}). 

As can be seen in the Fig. \ref{Re250}, $\epsilon^{1/2}$ for the  mode $m=0$ grows linearly with $\Omega$ with a characteristic change of slope occurring already around $\Omega=0.5$--$0.6$. The mode $m=1$ exists already at $\Omega=0$, which is typical for the \textit{2CRV} regime, but vanishes almost immediately as $\Omega$ increases and the axisymmetric flow prevails. The mode $m=1$ reappears at $\Omega = 0.5$--$0.6$ and its enstrophy $\epsilon$ grows linearly up to the value of $\Omega=1$, at which point pure \textit{low helical} regime first emerges. Simultaneously $m=0$ grows with $\Omega$ and the deviation from linearity is higher than that observed for lower $Re$. The phenomenon of decaying \textit{2CRV} regime, represented by mode $m=1$, and the subsequent appearance of axisymmetric flow (mode $m=0$) has been also numerically observed by \citet{JimenezJFM} in the study of a spinning bullet-shape body.

\begin{figure}

\hfill
\subfigure[$\epsilon^{1/2}$ of mode m=0 nondimensionalised with $D/v_{\infty}$]{
\label{allM0a}
\includegraphics[width=\plotsw pc]{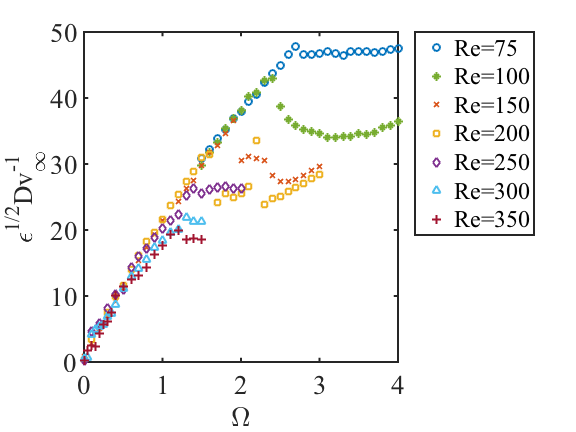}
}
\hfill
\subfigure[$\epsilon^{1/2}$ of mode m=0 nondimensionalised with $D/v_{\theta}$]{
\includegraphics[width=\plotsw pc]{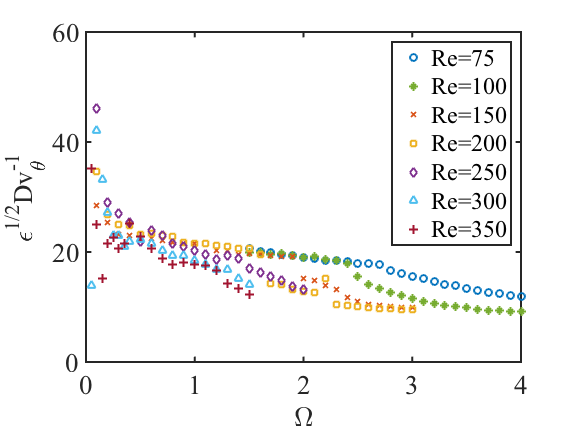}
\label{allM0b}
}
\caption{Dimensionless enstrophy of mode $m=0$ as a function of $\Omega$ and $Re$. In (b), where the enstrophy is undimensionalised with the vorticity induced by the rotation ($\sim v_{\theta}/D$), the initial mode $m=0$ is nearly constant and equal to 20. In the (a) the points of $Re=75$ are shifted, to maintain linearity, according to the $\epsilon_{plot}=\epsilon_{real}+\epsilon_{sh}$, where $\epsilon_{sh}=7.54$.}
\label{allM0}
\end{figure}

For all $Re$ the enstrophy of mode $m=0$ depends linearly on $\Omega$ up to a certain critical point beyond which it drops down and then seems to stagnate (see Fig. \ref{allM0a}). The value, of this critical $\Omega$, drops down with the increasing Reynolds number, i.e, $\Omega=2.6$ for $Re=75$ and $\Omega=0.6$ for $Re=350$. Taking $D/v_{\theta}$ as a dimensionless factor, the nearly constant value of enstrophy ($\epsilon^{1/2}Dv^{-1}_\theta\approx20$) of mode $m=0$ as a function of $\Omega$ can be observed (see Fig. \ref{allM0b}).

\subsection{Low helical state and mode $m=1$}
\begin{figure}

\hfill
\subfigure[mode m=0 of the average field]{
\label{Re150a}
\includegraphics[width=\plotsw pc]{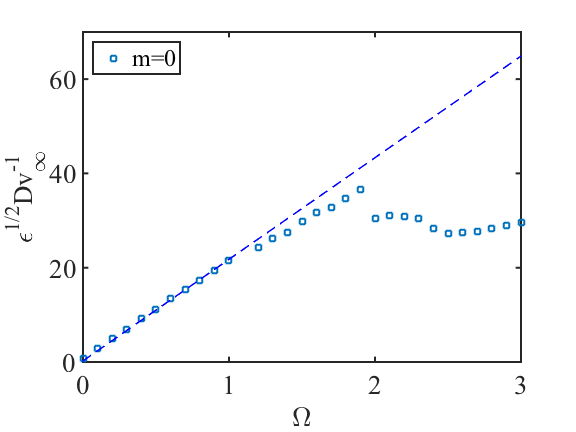}
}
\hfill
\subfigure[modes m=1,2,3,4]{
\includegraphics[width=\plotsw pc]{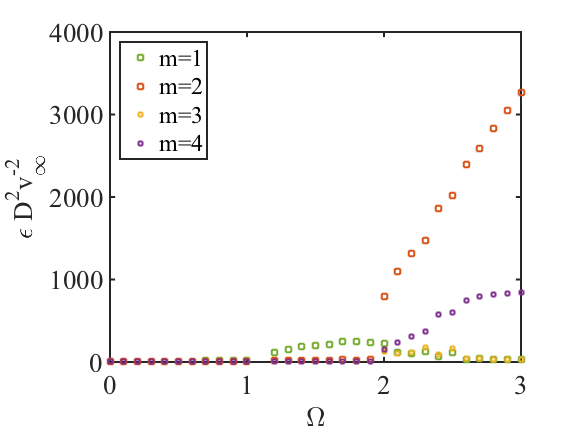}
}
\caption{Dimensionless enstrophy of modes as a function of $\Omega$, $Re=150$. (a) mode $m=0$ grows linearly up to $\Omega=1$, (axisymmetric flow), followed by small deviation due to \textit{LH}, and stagnation at $\Omega=2$ (\textit{HH}) (b) first detection of mode $m=1$ (\textit{LH}) at 1.2. The substantial increase of mode $m=2$ is at $\Omega=2$.}
\label{Re150}
\end{figure}

\begin{figure}
\hfill
\subfigure[Spectra for $Re=150$]{
\includegraphics[width=\plotsw pc]{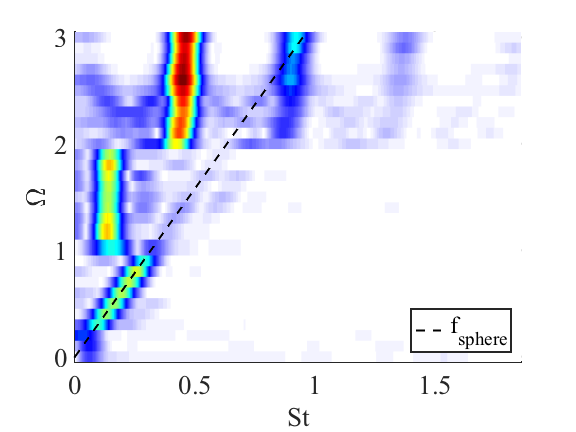}
\label{f150}
}
\hfill
\subfigure[Spectra for $Re=200$]{
\includegraphics[width=\plotsw pc]{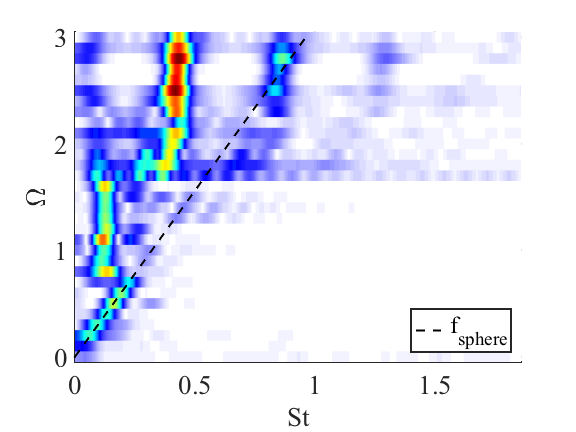}
\label{f200}
}
\caption{Spectra for $Re=150$ and $200$. Red and white denote the highest and zero amplitude, respectively. As $\Omega$ increases, three main groups was observed: linear growth with rotation frequency (dashed line), \textit{LH} frequency $St^1_{150}=0.14$ and $St^1_{200}=0.12$, \textit{HH} frequency $St^2_{150}=0.45$ and $St^2_{200}=0.41$}
\label{f150-200}
\end{figure}

The characteristic pattern of the \textit{low helical} regime was presented in the visualization for $\Omega=0.5$ (Fig. \ref{vis250}) and in the spatial-temporal reconstructions of PIV measurements for  $\Omega=0.5$ (Fig. \ref{piv}) as well as in the Table \ref{tabela} with the subtracted mode $m=0$.
Taking other modes into consideration we notice that mode $m=1$ appears for the first time for $Re=150$ (Fig. \ref{Re150}). This takes place for $1.0<\Omega<2.0$, for which the \textit{low helical} regime is observed and is evidenced both in the time-space reconstruction as well as in the visualization. The bifurcation point of the mode $m=2$ appears at $\Omega=2.0$. The case of $Re=150$ is therefore different than the lower Re cases, as the growth of the mode $m=0$ with $\Omega$ maintains linearity only up to $\Omega=1.1$ when the mode $m=1$ first appears. Then for $1.1<\Omega<1.9$ the slope of the linear trend is modified as the energy of the mode $m=0$ is partly transferred to mode $m=1$. For higher $\Omega$ the energy of the mode $m=0$ drops down and then stagnates. For $Re=200$ the mode $m=1$ appears first at $\Omega=0.5$ and then at $\Omega=1.5$ starts to decline, which coincides with the
strong burst of the mode $m=2$. It can be observed in Figs. \ref{f150} and \ref{f200} that in the \textit{low helical} regime the characteristic observed frequency is $St^1_{150}=0.14$ and $St^1_{200}=0.12$, where lower index stands for the Reynolds number, while the upper index (1) denotes \textit{low helical} regime. 

\begin{figure}\includegraphics[width=20 pc]{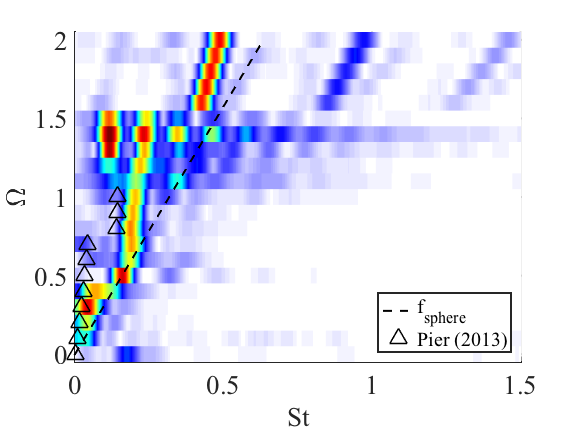}
\caption{Spectra for $Re=250$ and the superimposed results of \citet{Pier} related to the frequency of the leading eigenmode for axisymmetric basic wake. Red and white denote the highest and zero amplitude, respectively.}
\label{f250}
\end{figure}

For the case of $Re=250$ the \textit{low helical} regime appears when mode $m=1$ increases at $\Omega=0.4$ after presence of axisymmetric flow (see Fig. \ref{Re250}). In this regime (see Fig. \ref{f250}) for $\Omega=0.5$ to $1.1$, almost constant frequency ($St^1_{250}\approx0.21$) is visible in the spectrum. Additionally, for comparison with numerical study we superimpose to our results the numerical results from \citet{Pier} (Fig. \ref{f250}). This numerical study shows the frequency of approximately $^1/_3$ of $f_{rot}$ for $\Omega\leq0.7$ which only matches two of our experimental points i.e., $\Omega=0.1$ and $0.2$. Other values of frequency, nearly equal to 0.14, are in the different branch on higher Strouhal number for $0.8\leq\Omega\leq1$. These three points recover our branch at $St^1_{250}=0.21$.

For $1<\Omega<1.5$ and $Re=250$ the mode $m=1$ behaves erratically (see Fig. \ref{Re250}), while the enstrophy of mode $m=2$ becomes non-zero and exhibits moderate growth. In this range (see Fig. \ref{f250}), multiple frequencies are detected in the wake, including the frequency almost equal to $St^1_{250}$. Above $\Omega=1.5$ the enstrophy of the mode $m=0$ stagnates, the mode $m=1$ practically vanishes, while the modes $m=2$ and $4$ grow significantly - this corresponds to the presence of \textit{high helical} regime. 

\begin{figure}

\hfill
\subfigure[mode $m=1$]{
\label{allM1}
\includegraphics[width=\plotsw pc]{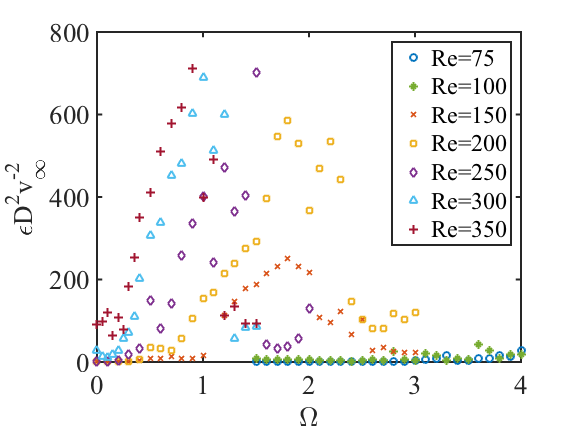}
}
\hfill
\subfigure[mode $m=2$]{
\includegraphics[width=\plotsw pc]{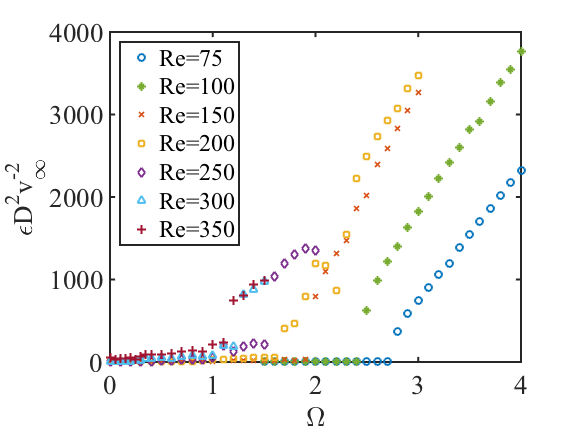}
}
\caption{Dimensionless enstrophy of mode $m=1,2$ as a function of $\Omega$ for all investigated $Re$. (a) bifurcation points for \textit{low helical} regime: starts with 0.25 for $Re=350$ up to 1.2 for $Re=150$ (b) bifurcation points for \textit{high helical} regime: 1.1 to 2.7 for $Re=350$ to $Re=75$, respectively.}
\label{allM2}
\end{figure}

The dimensionless enstrophy of mode $m=1$ as function of $\Omega$ for all Reynolds numbers is presented in Fig. \ref{allM1}. We observe that the nearly zero value is maintained for both $Re=75$ and $Re=100$. For $Re\geq150$ the mode $m=1$ is at first nearly zero, then increases rapidly, reaches the maximum, and then decays again to a zero value. The point where the growth appears is different for each $Re$, however the lowest value is reached for the highest $Re$, i.e., $\Omega=0.1$ for $Re=350$.

\subsection{High helical state and mode $m=2$}
\textit{High helical} regime can be characterized by patterns seen in the visualization for $\Omega=1.2$ (Fig. \ref{vis250}) and in the spatio-temporal reconstructions of PIV measurements for  $\Omega=1.2$ in Fig. \ref{piv}. In this regime, the wake consists of two pairs of helical vortices (a double helix) - each pair consisting of one positive and one negative longitudinal vortex (Tab. \ref{tabela} with the hidden mode $m=0$), which well agrees with visualizations of Section \ref{sectionVis}.

The \textit{high helical} regime is connected with the distinct appearance of the mode $m=2$. This distinguishes it from a \textit{low helical} regime, which is connected with the appearance of the mode $m=1$. The growth of mode $m=2$ is always related to the stagnation of mode $m=0$. Only for $Re=75$ and $Re=100$ the \textit{high helical} mode appears directly after the axisymmetric flow, for higher $Re$ the mode $m=2$ appears always preceded by mode $m=1$. This seemingly surprising sequence of modes was not observed here for the first time. In the study of swirling jets \cite{Meliga,Ruith} for certain values of the swirl and Reynolds number a subcritical transition from axisymmetric state to mode $m=2$ was identified.

For other, higher values of $Re$ the disappearance of mode $m=1$ leads to the strengthening of mode $m=2$ (Fig. \ref{allM2}). This bifurcation point for mode $m=2$ is different for each Reynolds number, but this is consistent with the behavior of modes $m=0$ and $1$. The rapid growth of mode $m=2$ is observed as follows: $\Omega=$ 2.7, 2.4, 1.9, 1.6, 1.5, 1.2, 1.1 for $Re=$75, 100, 150, 200, 250, 300, and 350 respectively. The mode $m=2$ of \textit{high helical} flow remains the most dominant mode for the highest $\Omega$ for all considered Reynolds numbers.

For this type of flow a strong characteristic frequency appearing in the wake is nearly constant for lowest Re, i.e., $St^2_{75}=0.47$ and $St^2_{100}=0.48$. These frequencies are accompanied by higher harmonics of the main frequency as visible in Figs. \ref{f75-100}(a-b). As can be seen in the Figs. \ref{f150-200}(a-b) and \ref{f250} the characteristic frequency of the \textit{high helical} mode slightly decreases with increasing $Re$, i.e., for $Re=150$ it is $St^2_{150}=0.45$ whereas for $Re=200$ it is $St^2_{200}=0.41$.

\begin{figure}

\centerline{\includegraphics[width=\plotsw pc]{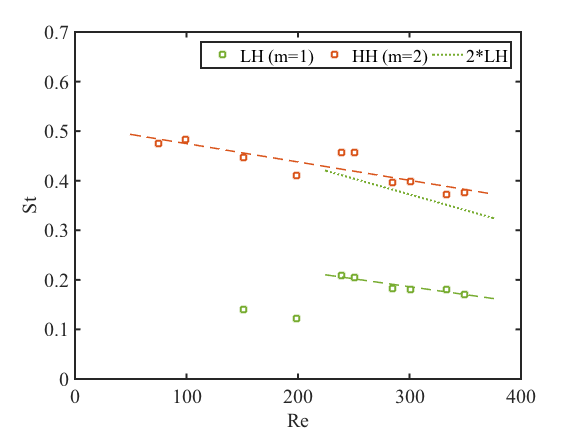}}
\caption{Main frequency as a function of $Re$. The frequency of \textit{HH} is nearly two times higher than frequency of \textit{LH}.}
\label{allFreq}

\end{figure}

The cases of $Re=$ 250, 300 and 350 are shown in the Figs. \ref{f250} and \ref{f300-350}(a-b). For these cases the frequency also decreases with increasing $Re$ but marginally increases with $\Omega$. An average value of this frequency was measured to be $St^2_{250}\approx 0.46$ and it is about two times higher than frequency $St^1_{250}$. For $Re=300$ and $350$ it is $St^2_{300}\approx 0.40$ and $St^2_{350}\approx 0.37$. All observed values of frequencies were also presented in Fig. \ref{allFreq}, showing that the Strouhal numbers of the \textit{high helical} mode is nearly two times the $St$ number of the \textit{low helical} mode $m=1$.

\begin{figure}

\hfill
\subfigure[mode $m=2$]{
\includegraphics[width=\plotsw pc]{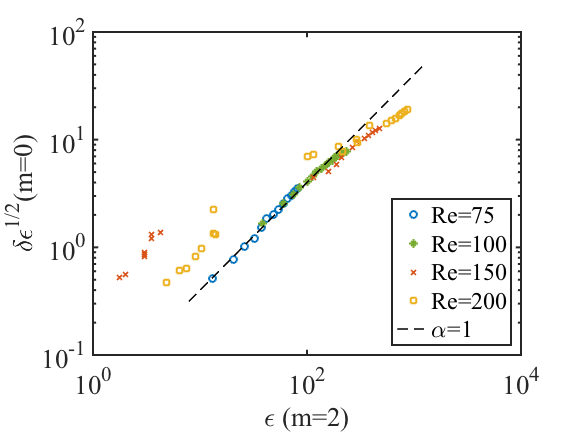}
\label{deltaM2}
}
\hfill
\subfigure[mode $m=4$]{
\includegraphics[width=\plotsw pc]{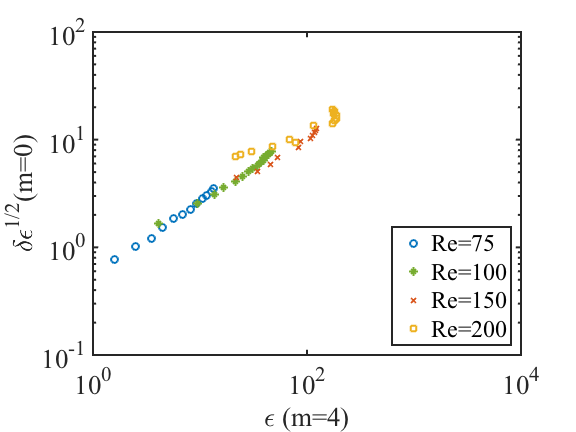}
\label{deltaM4}
}
\caption{$\delta\epsilon^{1/2}(m=0)$ as a function of enstrophy of mode $m=2$ and $m=4$ for different $Re$ in log-log scale. (a) the mode $m=2$ is proportional with coefficient 1 to the deviation from linearity of $m=0$, (b) the analogous proportionality coefficient for $m=4$ is nearly equal to 0.5}
\label{delta}
\end{figure}

In Figs. \ref{Re75} to \ref{Re350} the contribution of each mode to the entire flow was shown separately. The modes, however, are undoubtedly related and therefore it is worth to demonstrate this fact by representing one mode as a function of the other. Instead, however, of presenting the absolute value of enstrophy of mode $m=0$ we present its deviation from linearity, i.e., $\delta\epsilon^{1/2}(\Omega)=|\epsilon^{1/2}_{linear}(\Omega)-\epsilon^{1/2}_{real}(\Omega)|$, where $\epsilon^{1/2}_{linear}$ is a linear approximation of square-root of enstrophy of mode $m=0$ based on the lowest values of $\Omega$ and shown as a dashed line; the $\epsilon_{real}$ is a real value of enstrophy of modes.  The nearly zero value of $\delta$ indicates the linear behavior of the mode, while higher values of $\delta$ correspond to the significant deviation from linearity and this case is of particular interest.

In Fig. \ref{deltaM2} $\delta\epsilon^{1/2}$ is shown as a function of enstrophy of mode $m=2$ in the log-log scale. The mode $m=2$ persists for all investigated Reynolds numbers and for $Re\leq200$, a clear tendency to linear correlation has been observed with proportionality coefficient $\alpha=1$. The following relation can be written: $\delta m_0^\frac{1}{2} \propto m_2^1 $, where $m_i$ stands for the enstrophy of mode $i$.

In addition Fig. \ref{deltaM4} presents $\delta$ as a function of enstrophy of mode $m_4$ (log-log scale). The similar behavior to the case of mode $m=2$ is observed. Points corresponding to $Re\leq200$ are placed along one line with the proportionality coefficient higher than $0.5$. As the mode $m=1$ is not always observed it is difficult to regard this correlation as universal.

The change in slope of $\epsilon^{1/2}$ for the mode $m=0$, which grows linearly with $\Omega$ suggests the existence of a quadratic mode (``zero mode'') of modification of the basic state, as in 2D wakes \cite{EduardoZielinska}. Indeed the fluctuating modes (especially $m=2$) generate harmonics in frequency but activate also the homogeneous, steady ``zero mode'' related to the non-linearity of the flow dynamics. It is the first time when this kind of new behavior was observed for instabilities for the rotating bodies.

\subsection{Hairpin shedding}

\begin{figure}
\hfill
\subfigure[mode $m=0$ of the average field]{
\label{Re300a}
\includegraphics[width=\plotsw pc]{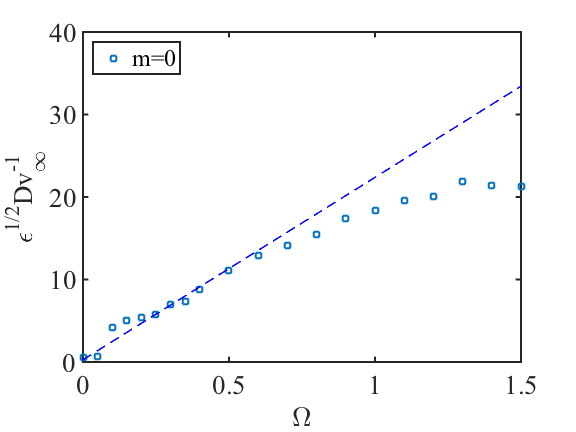}
}
\hfill
\subfigure[modes $m=1,2,3,4$]{
\includegraphics[width=\plotsw pc]{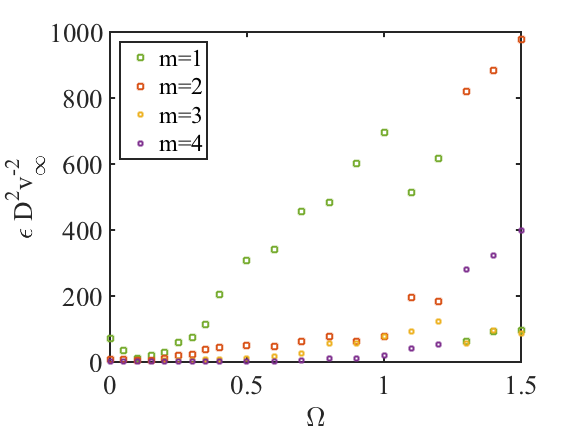}
}
\caption{Dimensionless enstrophy of modes as a function of $\Omega$, $Re=300$. (a) $m=0$ grows linearly up to $\Omega=0.6$ and stagnates at $\Omega=1.3$ (\textit{HH}), (b) an isolated point of $m=1$ of \textit{HS} is seen at $\Omega=0$, drops down immediately almost to 0 at $\Omega=0.1$--$0.2$, subsequently \textit{LH} is observed and $m=1$ continues to grow up to $\Omega=1.2$, when \textit{HH} appears.} 
\label{Re300}
\end{figure}

\begin{figure}
\hfill
\subfigure[Spectra for $Re=300$]{
\includegraphics[width=\plotsw pc]{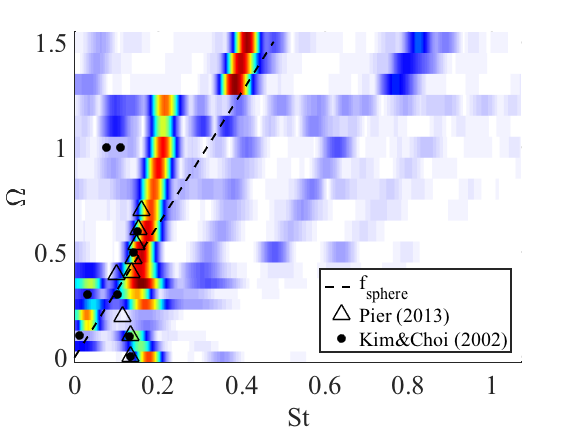}
\label{f300}
}
\hfill
\subfigure[Spectra for $Re=350$]{
\includegraphics[width=\plotsw pc]{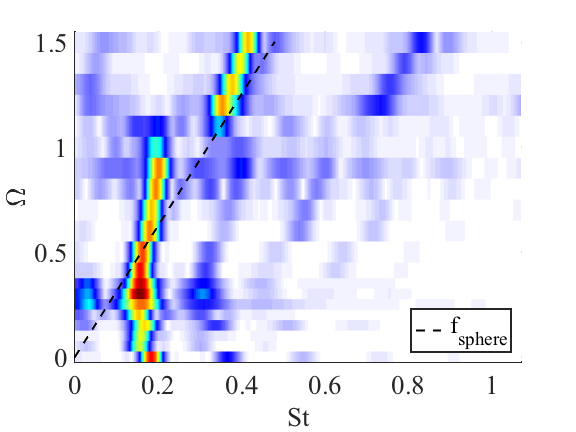}
\label{f350}
}
\caption{Spectra for $Re=300$ and $350$. Numerical results of \cite{KimChoi,Pier} were superimposed for $Re=300$. Strong correlation  between numerics and experiment can be seen. Red and white denote the highest and the zero amplitude, respectively}
\label{f300-350}
\end{figure}

\begin{figure}
\hfill
\subfigure[mode $m=0$ of the average field]{
\label{Re350a}
\includegraphics[width=\plotsw pc]{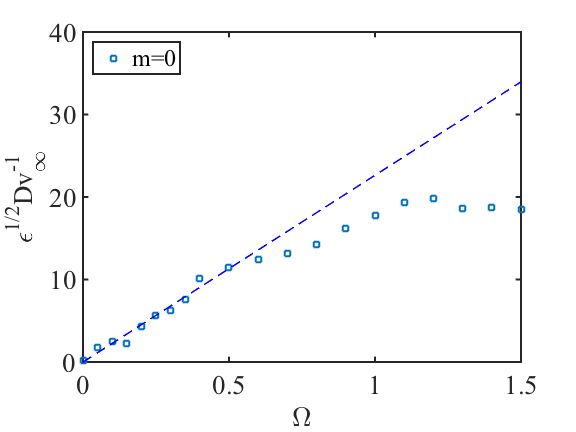}
}
\hfill
\subfigure[modes $m=1,2,3,4$]{
\includegraphics[width=\plotsw pc]{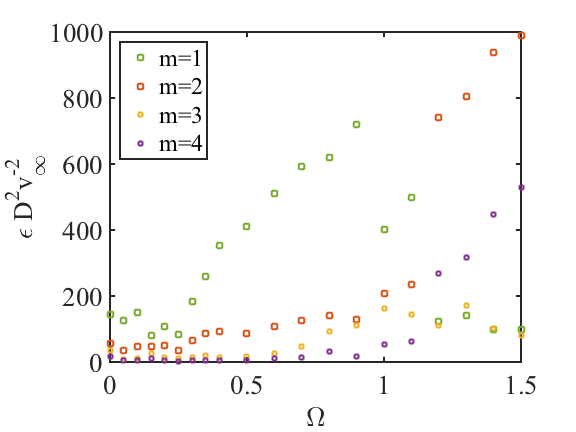}
}
\caption{Dimensionless enstrophy of modes as a function of $\Omega$, $Re=350$. (a) $m=0$ grows linearly up to $\Omega=0.5$ and stagnates at $\Omega=1.2$ (\textit{HH}), (b) $m=1$ of \textit{HS} is seen at $\Omega=0$, drops down and continues to grow while \textit{LH} is observed.}
\label{Re350}
\end{figure}

The flow scenario becomes very different for the case of $Re=300$ (Figs. \ref{Re300} and \ref{f300}) as this is the range above the second bifurcation of the static case ($Re_2=272$). Already for $\Omega=0$ we observe a strong frequency about $St^{HS}=0.18$, which corresponds to the hairpin shedding (\textit{HS}) regime observed earlier by \citet{Gumowski}. The amplitude corresponding to this frequency drops down and vanishes already around $\Omega=0.15$. The drop of the frequency has been also observed by numerical studies of \citet{KimChoi}, and \citet{Pier}. The positive correlation can be clearly seen in the Fig. \ref{f300}. For $\Omega=0.2$--$0.25$ the main observed frequency is significantly different from $St^{HS}$ and is lower than the frequency of rotation. A similar frequency has been observed by \citet{KimChoi}. The weakening of nearly \textit{HS} frequency confirms our earlier observation that the periodic \textit{HS} regime disappears as an effect of rotation. A similar behavior was observed by \citet{JimenezJFM} in the study of spinning bullet-shape body in a narrow region of $Re_2$. However from $\Omega=0.3$ the observed frequency ($St^1_{300}$) recovers and is only slightly larger than $St^{HS}$. 

This trend continues until $\Omega=1.2$ corresponding to the existence of the \textit{low helical} regime. In this range, we have quite a good agreement with the numerical results of \cite{KimChoi,Pier} with respect to the frequency. For $\Omega=1$ two frequencies have been observed by \citet{KimChoi} and we have also observed a similar weak frequency in this region for $\Omega=1.1$--$1.2$. Finally, for $\Omega=1.3$ to $1.5$ the characteristic frequency of the \textit{high helical} regime is observed $St^2_{300}=0.38$--$0.41\approx2St^1_{300}$.

\begin{figure}
\centerline{\includegraphics[width=16 pc]{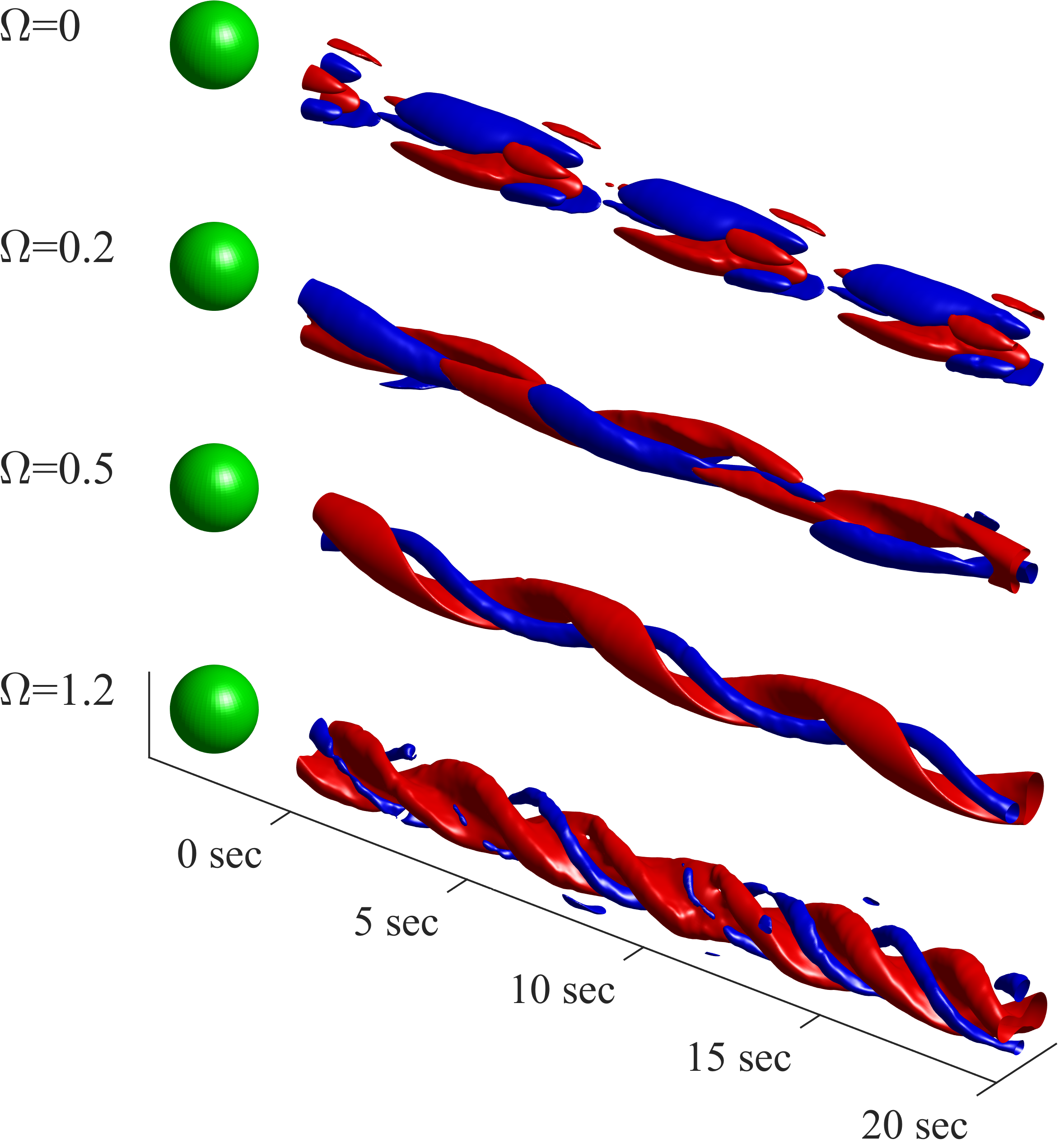}}
\caption{Spatio-temporal representation of the streamwise vorticity measured at $x=2D$ behind the sphere. $Re=350$, $\Omega$=0 (\textit{HS}), 0.2, 0.5 (\textit{LH}), 1.2 (\textit{HH}) from top to bottom. The red and blue isosurface corresponds to $\omega_x=\pm0.3\omega_{max}$ respectively, where $\omega_{max}$ denotes the maximum value of vorticity for a given case}
\label{rec350}
\end{figure}

The last investigated example corresponded to $Re=350$. Similarly to the previous case for $\Omega=0$, we observe a strong hairpin shedding with characteristic frequency $St^{HS}=0.18$. For $\Omega\geq0.05$ the amplitude related to this frequency slightly drops but eventually fully recovers. This frequency ($St^1_{350}\approx St^{HS}$) slightly increases up to the value of $0.19$. A new frequency $St^2_{350}\approx2St^1_{350}$ is observed for $\Omega\geq1.1$ and this is related to the appearance of a new \textit{high helical} regime ($m=2$). This observation is confirmed by the mode analysis (Fig. \ref{Re350}) as well as by the time-space PIV reconstructions (Fig. \ref{rec350}).

Figure \ref{rec350} shows the spatio-temporal representation of the streamwise vorticity for $Re=350$ and $\Omega=0, 0.2, 0.5, 1.2$. The typical pattern for hairpin shedding (\textit{HS}) is observed for $\Omega=0$. However, for $\Omega=0.2$ the pattern similar to the hairpin shedding is still observed. In the spectral analysis the similar characteristic frequency for \textit{HS} is detected also for $\Omega=0.2$. To clarify this phenomenon \textit{Dynamic Mode Decomposition} (DMD) was carried out to the longitudinal vorticity field $\omega_x(y,z,t)$ for $Re=300$ and $Re=350$, for which initially \textit{HS} exists.

This is an effective method to obtain the temporal description of coherent features existing in the flow-field. The flowfield can be in principle generated either by numerical simulations or experimental measurements (Schmid \cite{Schmid}, Jovanovic \textit{et al.} \cite{SchmidPHF}). Extraction of dynamic modes connected with a particular frequency can provide essential information on the spatial structure of the corresponding flow feature. The different regimes that can appear in the flow, depending both on the rotation rate and the Reynolds numbers, are characterized by different multiple frequencies observed in the wake.

\begin{figure}[h]
\centerline{\includegraphics[width=30 pc]{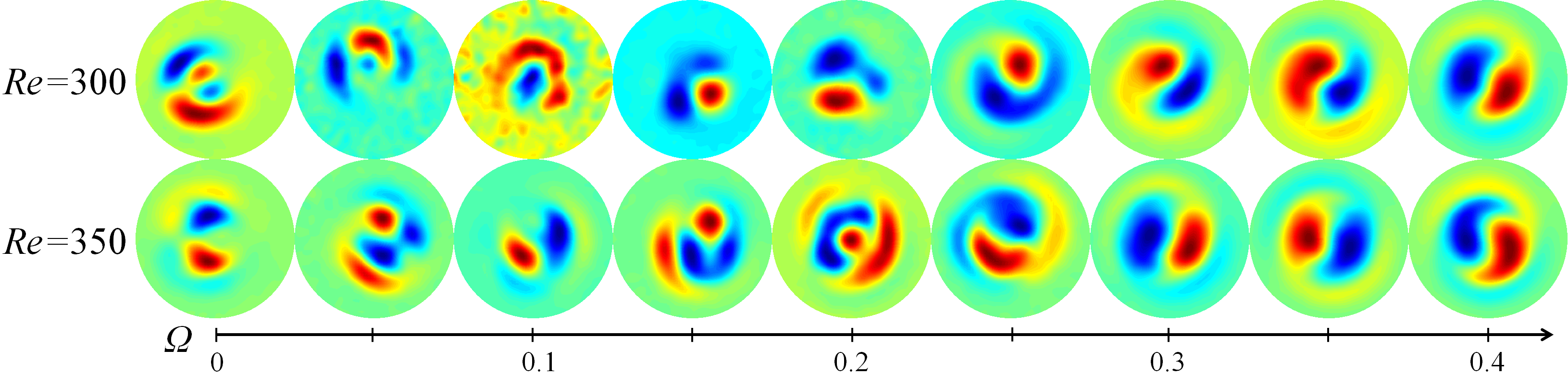}}
\caption{DMD modes related to the predominant frequency, \newline $Re=300$ (top) and $Re=350$ (bottom), $\Omega=0, 0.05, 0.1, 0.15 ... 0.4$ (left to right)}
\label{dmd}
\end{figure}

Figure \ref{dmd} presents spatial modes obtained by DMD for $Re=300$ and $Re=350$ and for $\Omega$ in the range $0$--$0.4$ with the step of $0.05$. These modes are relevant to the characteristic frequency described in this Section. For $\Omega=0$ the characteristic spatial pattern of \textit{HS} is observed for both Reynolds numbers, i.e., $Re=300$ and $Re=350$ as was observed in \cite{Szaltys,Bobinski,Chrust}. The similarities might be observed also for $\Omega=0.05$ however beyond this point the evolution of the flow is completely different. The hairpin shedding is no more observed for $Re=300$ for $\Omega>0.05$ which is consistent with the previous observations. For $Re=350$ the characteristic pattern of \textit{HS} is observed up to $\Omega=0.25$. Beyond this point, the spiral (or "Yin-Yang" helical) mode is observed for both considered $Re$ numbers. This kind of analysis clearly shows the evolution of initial hairpin shedding flow feature which is gradually decaying and finally is no longer observed for higher rotational rate $\Omega$. This effect has been also reported on the bifurcation map (Fig. \ref{map}). 

\section{Conclusions} 
The laminar flow past a sphere rotating in the streamwise direction was experimentally investigated. The research was performed for $Re=75$--$350$ and $\Omega$ in the range of $0$ and $4.0$. The investigation was carried out in a low-velocity water tunnel using LIF visualizations and PIV measurements. Different methods of data analysis were used to improve the clarity of results and make phenomena more understandable. The time-space reconstruction, the Azimuthal Fourier Decomposition, the Dynamic Mode Decomposition, as well as the frequency analysis were applied to the available data. The results show that the flow strongly depends on the rotation speed as well as on the initial regime (axisymmetric, \textit{2CRV} and \textit{HS}). The rotation of the sphere induces positive streamwise vorticity in the wake and this causes the gradual disappearance of the negative vortex and subsequently, the flow becomes axisymmetric. The further increase of the rotational rate causes the wake to become periodic and helical. Three new regimes for the rotating case were identified: axisymmetric flow or $m=0$ for $Re<250$ and various $\Omega$, \textit{low helical} or $m=1$ for $Re>150$ for moderate $\Omega=0.4$--$2$ and finally \textit{high helical} or mode $m=2$. This latter regime appears for all investigated Reynolds numbers for highest $\Omega$ typically $\Omega>1.5$.
We discovered also that the hairpin shedding instability is delayed by the rotation. It was shown for $Re=300$ that for the static sphere \textit{HS} appears. For very small $\Omega=0.1$ the periodic instability already disappears and the flow becomes steady one more time again. The case is different for larger Reynolds number when \textit{HS} instability is stronger. For $Re=350$ the \textit{HS} instability appears both for the static sphere as well as for $\Omega=0.1$. Moreover, the Dynamic Mode Decomposition analysis showed that the \textit{HS} instability persist longer for $Re=350$, namely for $\Omega=0$--$0.25$.  

Frequencies in the wake have been estimated for all the investigated cases. The appearance of new regimes is related to changes in the frequency spectrum. For the case of axisymmetric state a weak signal at the rotational frequency of the sphere was detected for $Re=75$, $Re=100$, $Re=150$, and $Re=200$. For the \textit{low helical} regime for both $Re=150$ and $Re=200$ a nearly constant frequency has been observed for $St^1=0.12$ to $0.14$. For higher $Re$ the measured frequency related to \textit{LH} is also nearly constant but higher than the former case ($St^1=0.2$) and it is slightly decreasing with the growing $Re$. Considering \textit{HH} state a different frequency appears. A very strong signal has been detected for all investigated $Re$ numbers. For the lowest $Re$, i.e. $75$ and $100$ the frequency related to the \textit{HH} is almost constant, however, the case is different for the highest $Re$ when the characteristic frequency is slightly increasing for higher rotational rates. The measured frequency is equal to $St^2=0.4$ to $0.5$ and it is slightly decreasing as for higher $Re$, similarly to the characteristic frequency of \textit{LH}. It has been also noticed that the \textit{HH} frequency is roughly two times higher than the \textit{LH} frequency. Finally, the frequency of the hairpin shedding has been detected for $Re=300$, $\Omega<0.15$ and for $Re=350$, $\Omega<0.25$ but in the latter case the frequency is difficult to distinguish from the \textit{LH} characteristic frequency. In both cases, the measured frequency of the \textit{HS} is equal to $St^{HS}=0.18$, which is in a good agreement with the previous experiments without rotation \cite{Gumowski}. Information about all observed regimes (axisymmetric, \textit{low helical} and \textit{high helical}) has been summarized in a bifurcation map as a function of both control parameters ($Re$ and $\Omega$). During the review process of this paper, a new numerical simulation of this problem has been conducted by \citet{JimenezNEW} for $Re=250$. They show a similar behavior to what we observed here, especially the value of transition from $m=1$ to \textit{high helical} mode $m=2$ was numerically predicted to be $\Omega=1.6$ which is exactly the same as reported here.

\begin{acknowledgments}
Special thanks to Beno\^{i}t Pier, Jos\'{e} Ignacio Jim\'{e}nez Gonz\'{a}lez, Marek Morzy\'{n}ski and Kornel Gibi\'{n}ski for fruitful discussions, and finally to Xavier Benoit Gonin for the technical support.
\end{acknowledgments}

\bibliographystyle{apsrev4-1}

\bibliography{mskarysz_PHF_v07}

\end{document}